\documentclass[twocolumn]{aastex631}

\usepackage{color}
\usepackage{amsmath}
\usepackage{graphicx}
\usepackage{graphics}
\usepackage{epstopdf}

\newcommand{\Mjup}{\mbox{$M_\mathrm{Jup}$}}
\newcommand{\Rjup}{\mbox{$R_\mathrm{Jup}$}}
\newcommand{\Msun}{\mbox{$M_{\odot}$}}

\shorttitle{H$\alpha$ Variability of AB Aur b with Accretion Light Echoes}
\shortauthors{Bowler et al.}
\graphicspath{{./}{figures/}}

\begin{document}

\title{H$\alpha$ Variability of AB Aur b with the Hubble Space Telescope: \\ Probing the Nature of a Protoplanet Candidate with Accretion Light Echoes}

\correspondingauthor{Brendan P. Bowler}
\email{bpbowler@astro.as.utexas.edu}

\author[0000-0003-2649-2288]{Brendan P. Bowler}
\affiliation{Department of Physics, University of California, Santa Barbara, Santa Barbara, CA 93106, USA}
\affiliation{Department of Astronomy, The University of Texas at Austin, Austin, TX 78712, USA}

\author[0000-0003-2969-6040]{Yifan Zhou}
\affiliation{Department of Astronomy, University of Virginia, 530 McCormick Rd, Charlottesville, VA 22904, USA}

\author[0000-0003-2646-3727]{Lauren I. Biddle}
\affiliation{Department of Astronomy, The University of Texas at Austin, Austin, TX 78712, USA}

\author[0000-0003-4006-102X]{Lillian Yushu Jiang}
\affiliation{Department of Astronomy, The University of Texas at Austin, Austin, TX 78712, USA}

\author[0000-0001-7258-770X]{Jaehan Bae}
\affiliation{Department of Astronomy, University of Florida, Gainesville, FL 32611, USA}

\author[0000-0002-2167-8246]{Laird M. Close}
\affiliation{Department of Astronomy, University of Arizona, 933 N. Cherry Ave., Tucson, AZ 85718, USA}

\author[0000-0002-7821-0695]{Katherine B. Follette}
\affiliation{Amherst College, Department of Physics and Astronomy, USA}

\author[0000-0003-4557-414X]{Kyle Franson}
\altaffiliation{NSF Graduate Research Fellow}
\affiliation{Department of Astronomy, The University of Texas at Austin, Austin, TX 78712, USA}

\author[0000-0001-9811-568X]{Adam L. Kraus}
\affiliation{Department of Astronomy, The University of Texas at Austin, Austin, TX 78712, USA}

\author[0000-0002-1838-4757]{Aniket Sanghi}
\altaffiliation{NSF Graduate Research Fellow.}
\affiliation{Cahill Center for Astronomy and Astrophysics, California Institute of Technology, 1200 E. California Boulevard, MC 249-17, Pasadena, CA 91125, USA}

\author[0000-0001-7258-770X]{Quang Tran}
\altaffiliation{51 Pegasi b Fellow}
\affiliation{Department of Astronomy, Yale University, New Haven, CT 06511, USA}

\author[0000-0002-4479-8291]{Kimberly Ward-Duong}
\affiliation{Five College Astronomy Department, Amherst College, Amherst, MA 01002, USA}

\author[0000-0002-4392-1446]{Ya-Lin Wu}
\affiliation{Department of Physics, National Taiwan Normal University, Taipei 116, Taiwan}

\author[0000-0003-3616-6822]{Zhaohuan Zhu}
\affiliation{Department of Physics and Astronomy, University of Nevada, Las Vegas, 4505 South Maryland Parkway, Las Vegas, NV 89154-4002, USA}

\begin{abstract}
Giant planets generate accretion luminosity as they form. Much of this energy is radiated in strong H$\alpha$ line emission, which has motivated direct imaging surveys at optical wavelengths to search for accreting protoplanets. However, compact disk structures can mimic accreting planets 
by scattering emission from the host star. This can complicate the interpretation of H$\alpha$ point sources, 
especially if the host star itself is accreting. 
We describe an approach to distinguish accreting protoplanets from scattered-light disk features using ``accretion light echoes.'' 
This method relies on variable H$\alpha$ emission from a stochastically accreting host star 
to search for a delayed brightness correlation with a candidate protoplanet.
We apply this method to the candidate protoplanet AB Aur b with a dedicated Hubble Space Telescope Wide Field Camera 3 
program designed to sequentially sample the 
host star and the candidate planet in H$\alpha$ while accounting for the 
light travel time delay and orbital geometry of the source within the protoplanetary disk.
Across five epochs spanning 14 months, AB Aur b is over 20 times more variable than its host star; 
AB Aur's H$\alpha$ emission changes by 15\% while AB Aur b varies by 330\%.
These brightness changes are not correlated, 
which rules out unobstructed scattered starlight from the host star as the only source of AB Aur b's H$\alpha$ 
emission and is consistent with tracing emission from 
an independently accreting protoplanet,
inner disk shadowing effects, or a physically evolving compact disk structure.
More broadly, accretion light echoes offer a novel tool to explore the nature of
protoplanet candidates with well-timed observations of the host star prior to deep imaging in H$\alpha$.
\end{abstract}


\section{Introduction} \label{sec:intro}

Giant planets assemble the vast majority of their mass by accreting material from circumplanetary disks
nested within more massive gas-rich protoplanetary disks.  
This process of giant planet growth is not well understood compared to accretion in the stellar regime,
where infalling gas is heated as it accelerates along
magnetic field lines and collides with the protostar to produce shocks and hot spots (e.g., \citealt{Hartmann:2016gu}).  
Together these give rise to hot continuum emission in the UV, veiling across the optical spectrum,
and strong, broad, emission lines
(e.g., \citealt{Herbig:1962aa}, \citealt{Calvet:1998fu}; \citealt{Ingleby:2013hf}).

In the planetary regime, the geometry of accretion is still an open question and likely depends on the
planetary magnetic field strength and properties of the circumplanetary disk.
For example, gas accretion may proceed through spherically symmetric flows, predominantly at the poles,
or through magnetospheric accretion columns (\citealt{Zhu:2015fr}; \citealt{Aoyama:2018fg}; \citealt{Thanathibodee:2019kn}; \citealt{Marleau:2022bb}).
However, even the basic elements underpinning these models are largely unconstrained by observations
including the characteristic timescale and location of giant planet formation;
the behavior of planetary accretion rate histories (e.g., steady-state or episodic; \citealt{Fortney:2008ez}; \citealt{Dong:2021bb}); 
circumplanetary disk masses, temperatures, and geometries (\citealt{Ward:2010fc}; \citealt{Lubow:2012ea}; \citealt{Zhu:2016iu});  
and the mechanism of mass and angular momentum transfer from the protoplanetary disk through the
circumplanetary disk and onto the young planet (\citealt{Bryan:2018hd}; \citealt{Batygin:2018hq}; \citealt{Bryan:2020ex}).
Finding and confirming young accreting protoplanets is the first step to begin to address these questions.

Directly detecting UV continuum emission from young planets requires space-based observations (e.g., \citealt{Zhou:2021ky}),
but optically bright hydrogen recombination lines---most notably H$\alpha$---offer a convenient approach to search for accreting planets.
However, high-contrast imaging in H$\alpha$ has been difficult 
because of technical challenges and conflicting interpretations of results.
Ground-based adaptive optics (AO) systems are traditionally optimized for infrared wavelengths,
so specialized instruments are needed to achieve high Strehl ratios in the visible 
(e.g., MagAO, \citealt{Close:2014tn}; MagAO-X, \citealt{Males.2020aa}; SCExAO/VAMPIRES, \citealt{Norris:2015jw}; SPHERE/ZIMPOL, \citealt{Schmid:2018aa}; VLT/MUSE, \citealt{Bacon:2010aa}).  
This has enabled the first generation of high-contrast imaging surveys targeting 
young stars in H$\alpha$ (e.g., \citealt{Cugno:2019aa}; \citealt{Zurlo:2020mm}; \citealt{Xie:2020aa}; \citealt{Huelamo:2022aa}; \citealt{Follette:2023aa}; \citealt{Cugno:2023aa}).
Some space-based imaging in H$\alpha$ has been carried out with the Hubble Space Telescope,
but its modest angular resolution and lack of advanced coronagraphic capabilities for starlight suppression 
have limited surveys to modest sample sizes (e.g., \citealt{Zhou:2014ct}; \citealt{Sanghi:2022kn}; \citealt{Zhou:2022fa}).

Although several accreting protoplanet candidates were identified in these campaigns,
many of these detections and their interpretations have been contested.  
LkCa~15 is a prominent example; 
multiple point sources were identified  
in thermal emission and H$\alpha$ (\citealt{Kraus:2012gk}; \citealt{Sallum:2015ej}), but their nature has been 
called into question because of confusion with inner disk structures 
(e.g., \citealt{Thalmann:2016kd}; \citealt{Currie:2019aa}; \citealt{Blakely:2022aa}).
Compact features from an inner disk may be mimicking planets
by preferentially scattering light from the host star.  These bright regions can appear point-like and 
result in false positive planet signals (\citealt{Follette:2017jw}; \citealt{Rameau:2017js}).
\citet{Sallum:2023aa} found that the LkCa~15 inner disk substructure is variable and may be tied to
physical evolution of these features on timescales of months to years, illustrating the challenges
involved in validating young planet candidates.

The clearest case of accreting protoplanets is in the PDS 70 system,
which harbors two directly imaged planets nested within a large transition disk gap 
(\citealt{Keppler:2018dd}; \citealt{Muller:2018aa};
\citealt{Haffert:2019ba}).  Both planets were found to be accreting through deep H$\alpha$
imaging (\citealt{Wagner:2018bb}; \citealt{Haffert:2019ba}; \citealt{Hashimoto:2020gn}; \citealt{Zhou:2021ky}; \citealt{Close:2025aa}; Zhou et al. 2025, in press), 
and circumplanetary
dust has been detected around PDS 70 c with ALMA (\citealt{Isella:2019em}; \citealt{Benisty:2021ft}).  
This system has provided the clearest window into
the planet assembly process and the properties of their circumplanetary disks. 

Recently, a concentrated feature was detected within the protoplanetary disk surrounding AB Aurigae 
(AB Aur; \citealt{Currie:2022dd}; \citealt{Zhou:2022fa}), 
a young ($\approx$2--4~Myr), intermediate-mass ($\approx$2.4~$M_{\odot}$) accreting A0 star in the greater Taurus-Auriga complex (\citealt{DeWarf:2003aa}).
AB Aur and its large complex disk have been well-studied for many decades (e.g., \citealt{Sanford:1958aa}; \citealt{Marsh:1995aa}; \citealt{Mannings:1997aa}; \citealt{Betti:2022aa}).
Millimeter observations show a large dust cavity out to about 120 AU with two spiral structures detected within the dust ring (\citealt{Tang:2017gr}; \citealt{Jorquera:2022cs}).  A central compact radio-bright source has been found to be consistent with an outflow from AB Aur (\citealt{Rodriguez:2014aa}; \citealt{RiviereMarichalar:2023aa}; \citealt{Rota:2024aa}).
\citet{Grady:1999aa} identified regular sporadic infall of material onto AB Aur which may be related to evaporating exocomets,
and infrared variability of the disk itself has been observed over a broad range of timescales and wavelengths (\citealt{Shenavrin:2019aa}; \citealt{Prusti:1994aa}; \citealt{Chen:2003aa}).
On larger scales, outer spiral arms extend to at least $\sim$450 AU (\citealt{Grady:1999aa}; \citealt{Fukagawa:2004kp}; \citealt{Corder:2005aa}; \citealt{Perrin:2009fr}).
\citet{Speedie:2024aa} recently found evidence of gravitational instability from kinematic signatures in ALMA isovelocity maps, suggesting that the
disk may be susceptible to gravitational collapse.


\begin{figure*}
  \vskip -1.2 in
  \hskip -1.35 in
  \resizebox{10.5in}{!}{\includegraphics{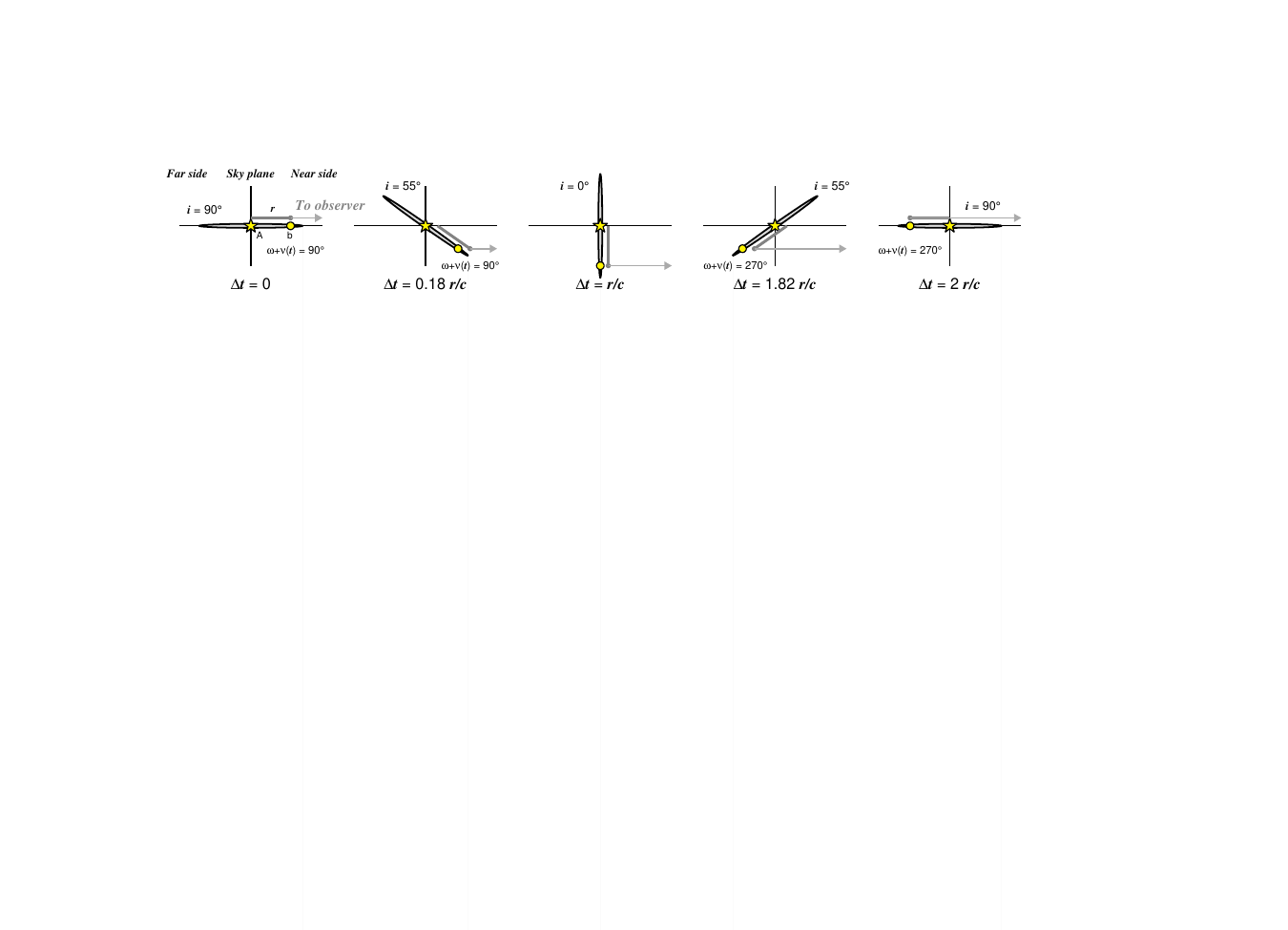}}
  \vskip -5.1 in
  \caption{Examples of the impact of orbital geometry on light echo time delays.  Stochastic changes in the brightness of a host star (from episodic accretion, for instance)
  will be seen as delayed variability from a compact disk feature at an orbital distance $r$.  This time delay will range from 0 at inferior conjunction 
  ($\omega + \nu(t)$ = $\pi$/2; left panel) 
  to $r/c$ in the sky plane (middle panel) to 2$r/c$ at superior conjunction ($\omega + \nu(t)$ = 3$\pi$/2; right panel).  \label{fig:geometry} } 
\end{figure*}

There have been hints of planets within the AB Aur disk in several datasets (e.g., \citealt{Oppenheimer:2008gn}; \citealt{Boccaletti:2020ba}),
including the faint source (``AB Aur b'') identified by \citet{Currie:2022dd} at a projected separation of $\approx$0$\farcs$6 ($\approx$90 AU). 
\citet{Currie:2022dd} recovered AB Aur b in multiple epochs from the ground and archival Hubble Space Telescope (HST)
observations spanning the optical to near-infrared.  They argued that the resolved nature of the source, the lack of a detection in polarized light, signs of orbital motion, 
and the spatial location
coincident with the trailing edge of an inner spiral arm all point to an embedded giant planet.  

\citet{Zhou:2022fa} identified the source in H$\alpha$ as part of a high-contrast imaging survey of transition disk hosts 
with HST's Wide Field Camera 3 (WFC3).
They found elevated emission but note that the H$\alpha$ line-to-continuum ratio is similar to that of the host star, raising the possibility 
that AB Aur b is a compact disk structure and the H$\alpha$ line emission is reflected light from the accreting host star.
Moreover, strong scattered light features are prevalent throughout the disk 
including in the vicinity of AB Aur b.
In a follow-up study, \cite{Zhou:2023di} recovered AB Aur b in the UV and blue optical with HST and confirmed that its short-wavelength spectral energy distribution (SED) 
is dominated by scattered stellar light based on a strong Balmer break, which would not be expected for a planet but is in good agreement
with reflected light from the A0 host star.  
Despite this progress, the true nature AB Aur b remains elusive and
more observational tests are needed to distinguish between a disk structure or an embedded accreting protoplanet.

Here, we present a new 
method to test whether an H$\alpha$ point source is an accreting planet or a compact disk feature seen in scattered light.
This strategy relies on ``accretion light echoes'' caused by variable accretion onto the host star 
to probe the nature of the companion.
In combination with other lines of evidence, this technique can be used distinguish accreting planets 
like those orbiting PDS 70 from false positive signals caused by disk substructure, as appears to be the case with LkCa~15.
We apply this technique to the recently identified candidate protoplanet AB Aur b 
with a dedicated narrow-band H$\alpha$ imaging program using HST/WFC3, in which
carefully timed sampling of the host star is carried out prior to deep observations of the companion
across five epochs to
establish whether stochastic accretion from the host star is also witnessed in variability of the companion.

This study is organized as follows.  In Section~\ref{sec:echoes} we describe the geometric setup of the accretion light echo problem
and discuss measurements that are needed to determine the orbital orientation of the planet candidate.
We present the HST observations of AB Aur in Section \ref{sec:obs} and
results of the accretion light echo experiment for AB Aur b in Section \ref{sec:results}.  
Potential interpretations of the variability results are discussed in Section \ref{sec:discussion}.
A summary of results can be found in Section \ref{sec:summary}.

\section{Accretion Light Echoes as a Test of the Protoplanet Hypothesis}{\label{sec:echoes}}

Light echoes have long been used as a tracer of dust structure in the vicinity of time-variable sources.  
Applications include the identification of historic supernovae (\citealt{Rest:2012aa}),  probing AGN accretion disks via reverberation mapping (\citealt{Bahcall:1972aa}; \citealt{Blandford:1982aa}), 
resolving structure in protoplanetary disks (\citealt{Gaidos:1994aa}; \citealt{Ortiz:2010aa}), and even searching for planetary companions
(\citealt{Argyle:1974aa}; \citealt{Bromley:1992aa}; \citealt{Sparks:2018aa}; \citealt{Bromley:2021aa}).
This technique relies
on variable emission from a central source and subsequent delayed scattering of that signal, usually from nearby circumstellar or interstellar dust.
Because scattering from a compact disk feature can mimic an accreting protoplanet, this method can naturally be used to test the
nature of H$\alpha$ point sources around young stars such as AB Aur in the absence of other sources of variability. 

\subsection{Geometry of Accretion Light Echoes}

The travel time for a pulse of light originating from the host star to a source at an orbital distance $r$ is $r$/$c$.
The time delay ($\Delta t$) for a distant observer to see the same signal scattered from the orbiting body depends on the orbital geometry
and its position in its orbit relative to the observer's line of sight.  This orientation can be characterized by $\omega$ + $\nu (t)$, where $\omega$ is argument of periastron---the
angle between the longitude of ascending node and periastron passage---and the true anomaly $\nu$---the time-dependent angle between periastron 
and the position of the orbiting body.  When the planet is on the near side of its orbit closest to Earth, $\omega$ + $\nu (t)$ spans 0--$\pi$, and on the far side of the orbit $\omega$ + $\nu (t)$ spans $\pi$--$2 \pi$.

In the context of using an accretion light echo to probe the nature of a directly imaged H$\alpha$ source,
the goal is to test whether the source is reflecting light from the host star or whether its brightness variations are behaving independently.
In the former case, 
any variable emission (e.g., H$\alpha$ line strength) from the companion 
would be caused by earlier stochastic changes in the accretion rate onto the host star.
Here, the orbit of the H$\alpha$ source can be assumed to lie within the plane of a protoplanetary disk.
Figure~\ref{fig:geometry} illustrates some simple examples for varying disk inclinations.
For an edge-on disk, a dust clump located at inferior conjunction (on the front side of the disk closest to the observer where $\omega$ + $\nu(t)$ = $\pi$/2), 
there is no lag between emission
from the host star and scattered light so $\Delta t$ = 0.  At superior conjunction (on the far side of the disk where $\omega$ + $\nu(t)$ = 3$\pi$/2), 
the distance traveled is twice the orbital radius and the maximum delay time is $\Delta t$ = 2$r/c$.
Positions in the plane of the sky produce a delay of $\Delta t$ = $r/c$, which occur for a face-on orientation, 
for the quadrature points of a circular orbit, or more generally when $\omega$ + $\nu(t)$ is equal to 0 or $\pi$.
For context, this characteristic timescale $r/c$ is 42~min at 5 AU and 13.9 hr for a distant companion at 100 AU.
All other orbital geometries will fall between these extremes (0--2 $r/c$) and 
will depend on the orbital inclination $i$ with respect to the plane of the sky and orbital phase in the following way:

\begin{equation}
\Delta t = \frac{r}{c}\Big( 1 - \sin(i) \sin(\omega + \nu(t)) \Big).
\end{equation}

\begin{figure}
  \hskip -0.2 in
  \resizebox{3.7in}{!}{\includegraphics{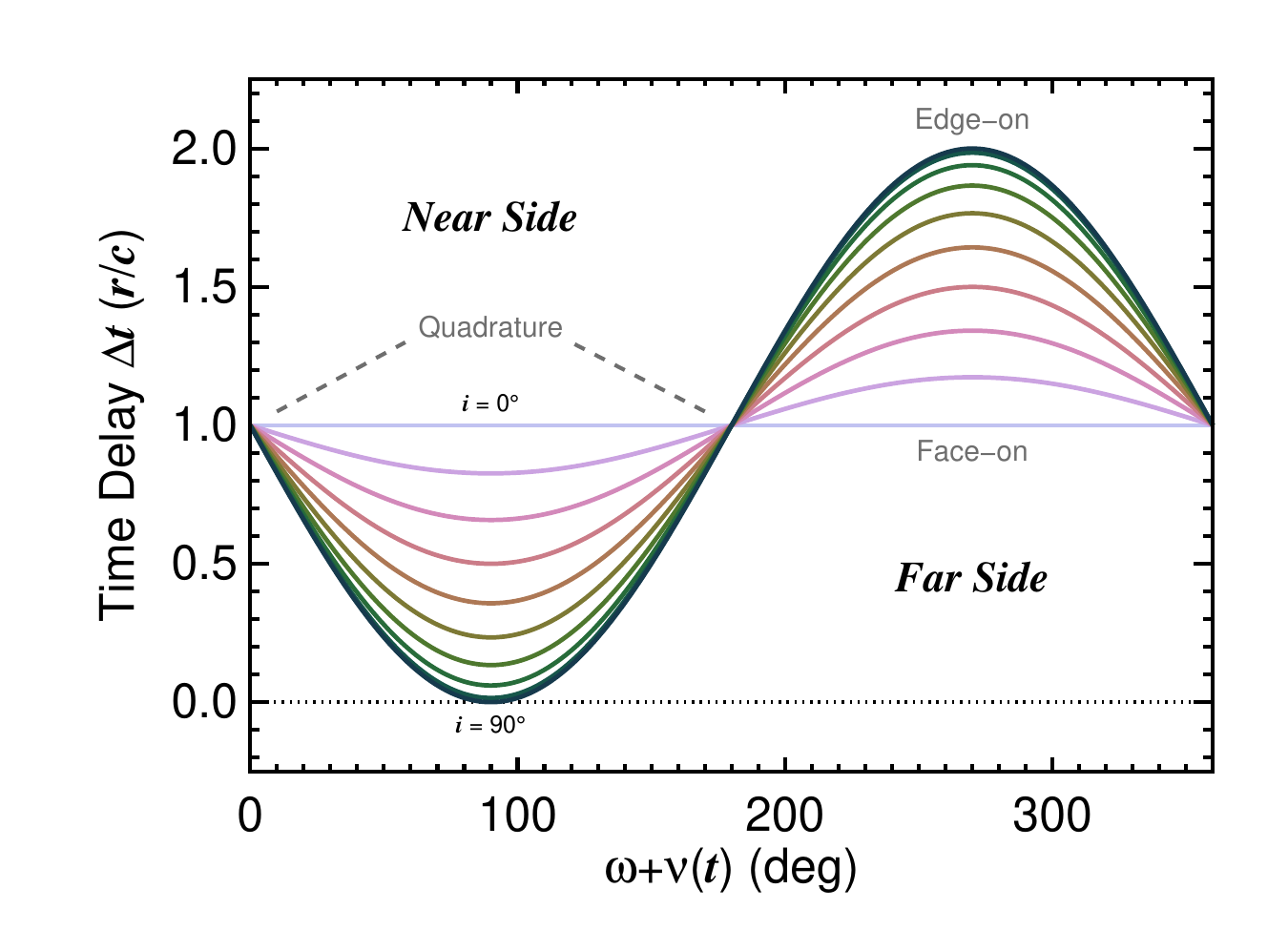}}
  \vskip -0.1 in
  \caption{Time delay of light echoes as a function of orbital phase ($\omega$ + $\nu(t)$).  Time delays
  are shown in units of $r/c$, which is the time it takes for light to travel from the star to the companion.  Inclinations from 0$\degr$ to 90$\degr$ are shown in steps of 10$\degr$. 
  Depending on the inclination and location in the orbit, the time delay can reach a minimum of 0 and a maximum of 2$r/c$,
  depending on the three-dimensional geometry of the system.  For the application 
  to accreting protoplanets, this highlights the need to determine the disk inclination, the disk orientation, 
  and the current position of the companion in its orbit to accurately compute 
  the expected time delay for a light echo.    \label{fig:general_delay} } 
\end{figure}

\noindent The first term ($r/c$) is the time it takes for light to travel from the host star to the companion at a distance $r$,
and the second term is the time to traverse the component of the companion's position along the line of sight orthogonal to the sky plane.
The difference is the extra time (and distance) to reach the observer after the scattering event.
This relationship assumes a circular orbit but can readily be extended to elliptical orbits
with a time-dependent orbital distance (e.g., \citealt{Sparks:2018aa}).
Figure~\ref{fig:general_delay} illustrates this time lag as a function of orbital phase for various orbital inclinations.


\subsection{Application to AB Aurigae}{\label{sec:abaurgeometry}}

This method of accretion light echoes requires that the host star be sufficiently variable, either in continuum or 
line emission (such as H$\alpha$ or Pa$\beta$).  Moreover, if the rate of stochastic brightness variations is 
slow---for instance, days or weeks---compared
to the light travel time of the orbiting companion, then the impact of the time delay will not be important.  In this case simultaneous
observations of the host and companion would be sufficient for this experiment.  On the other hand,
if episodic accretion is fast---on timescales of minutes or hours---then the time delay should properly be taken into account, and 
separate observations are needed for the host and then again later for the companion.
The variability properties of the host star are therefore relevant for the design of an accretion light echo experiment.

AB Aur is actively accreting at a rate of $\approx$10$^{-7}$ $M_{\odot}$ yr$^{-1}$ (\citealt{Salyk:2014hc}) 
and shows strong changes in the H$\alpha$ emission line profile and overall strength at the level of tens of percent 
on a variety of timescales ranging from hours to years 
(e.g., \citealt{Herbig:1960aa}; \citealt{Finkenzeller:1983aa}; \citealt{Catala:1999aa}; \citealt{Harrington:2007bq}; \citealt{Costigan:2014aa}).  
Moreover, the H$\alpha$ emission line intensity is much greater than the continuum level: \citet{Harrington:2007bq} find typical 
peak line intensities about 10 times higher than the continuum emission.  
Changes in the H$\alpha$ emission line strength are 
therefore expected dominate the 
overall variability at this wavelength.

\begin{figure}
  \resizebox{3.2in}{!}{\includegraphics{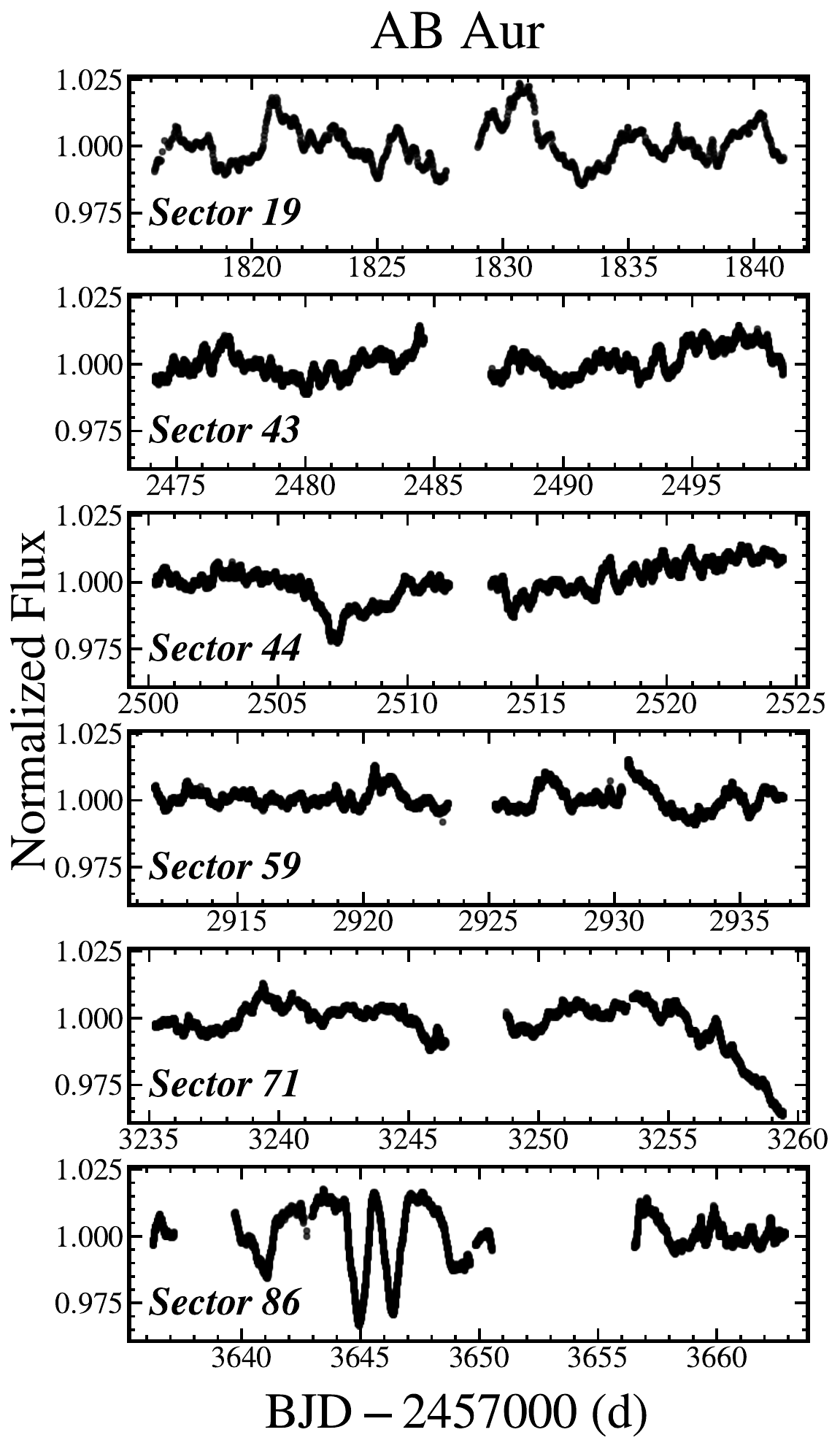}}
  \vskip -0.1 in
  \caption{Normalized TESS light curves of AB Aur from Sectors 19, 43, 44, 59, 71, and 86.
  The variability within each Sector of a few percent suggests only modest changes in accretion rate
  over month-long baselines.    \label{fig:tess} } 
\end{figure}

Accretion also produces significant broadband variability.  AB Aur shows brightness changes at the level of 0.1~mag
in optical light curves from the Microvariability and Oscillations of Stars (MOST) satellite spanning 24 days in 2009 and 2010
(\citealt{Cody:2013ek}).  More recently, AB Aur was monitored in six 27-day sectors from December 2019 to December 2024 by the 
Transiting Exoplanet Survey Satellite (TESS; \citealt{Ricker:2014ua}); details of the TESS light curve extraction and analysis are discussed in Section~\ref{sec:tess}.
The TESS lightcurves show short-term brightness changes at the few-percent level  
over day-to-week timescales (Figure~\ref{fig:tess}). 
Interestingly, the behavior of AB Aur in the latest TESS lightcurve (Sector 86 in December 2024) shows sharp 3\% variations that fall and
rise on day-long timescales, which bear some resemble to dipper stars (albeit with lower amplitudes) and could be a sign that
close-in occulting circumstellar disk material is contributing to the optical variations (e.g., \citealt{Capistrant:2022ci}).
Altogether the combination of rapid accretion-induced broadband and H$\alpha$ variability make AB Aur an excellent testbed for
carrying out an accretion light echo experiment with the wide candidate protoplanet AB Aur b.


For most applications of accretion light echoes, the orbit of the source of interest may not be well constrained if it was recently 
discovered or if the candidate is located at a wide separation with a correspondingly long orbital period.
Moreover, even with full coverage of the orbital period, an orbit based on relative astrometry 
alone is not enough to distinguish the full three-dimensional geometry of the orbital plane
because there exists a 180$\degr$ degeneracy in the longitude of ascending node.  
This is the case for AB Aur b, which orbits due south of its host star at a projected separation of 0$\farcs$6, or 93~AU.  
If AB Aur b is on the near side of the sky plane, the light echo time delay will be less than $r/c$,
but if it is on the far side, the lag time greater than $r/c$.  
There are ways to break this degeneracy using disk kinematics or spiral arm morphology, as discussed below.

\citet{Currie:2022dd} detected orbital motion in the counterclockwise direction based on over a decade
of astrometric measurements.\footnote{Additional epochs obtained with HST by \citet{Zhou:2022fa}
and \citet{Zhou:2023di} do not appear to follow the same clear trend.  This could be caused by instrument-to-instrument systematics
or perhaps alternative approaches to modeling the position of the source given its resolved nature.
Nevertheless, for this study we treat counterclockwise motion as a true indication of the orbital direction of AB Aur b.}
Information about the geometric orientation of the orbit can be obtained from
spatially resolved mm imaging of the disk to yield its inclination and probe its velocity structure.
Together this is enough to break the degeneracy in the longitude of ascending node and determine
if the southern side of the inclined disk is closer to Earth than the northern side, or vice versa.

There is a a long history of spatially resolved mapping of AB Aur's dust and gas disk spanning the near-infrared to cm wavelengths.
Although the physical properties of the disk have been studied in depth, the large size and complex morphology of the disk 
have at times yielded disk parameters that disagree in their details.
Below we summarize recent observational constraints on the inclination and geometric orientation of AB Aur's disk, which we then
use to determine the expected delay time for an accretion light echo experiment.

Scattered-light $H$-band imaging from \citet{Fukagawa:2004kp} yielded $i_d$ = 30 $\pm$ 5$\degr$ and a P.A. of 58 $\pm$ 5$\degr$
for the orientation of the major axis of the disk.
Similarly, \citet{Perrin:2009fr} found a range of 22--35$\degr$ for the disk inclination 
from HST/NICMOS scattered-light polarized and total-intensity observations,
and \citet{Hashimoto:2011bt} determined an inclination of 27--43$\degr$.

Resolved millimeter studies offer an alternative approach to measuring the disk orientation, rotation axis, and gas kinematics.
At a scale of $\sim$100 AU, \citet{RiviereMarichalar:2020aa} found an inclination of $\approx$24.9$\degr$ and a P.A. of the disk axis of $\approx$--37$\degr$ based on  1.3~mm NOrthern Extended Millimeter Array (NOEMA) interferometric observations.
\citet{Tang:2017gr} reported high-resolution (0$\farcs$1) ALMA 1.3~mm continuum and CO line observations of the 
dust ring and inner CO spiral structure.  They found a disk inclination of 23$\degr$ and a P.A. of --36$\degr$ for the disk rotation axis.
More recently, \citet{Speedie:2024aa} observed AB Aur with ALMA in Band 6 and found a P.A. of 236.7 $\pm$ 0.3$\degr$ 
(measured east of north to the redshifted major axis) from fitting the C$^{18}$O moment1 map and adopting a disk inclination of 23$\degr$.

Complex structure has been observed in the AB Aur disk at larger spatial scales of several hundred AU. 
\citet{Corder:2005aa} observed AB Aur at 2$''$ resolution 
in $^{13}$CO, $^{18}$CO, $^{12}$CO, and 2.7 mm continuum emission with the Owens Valley Radio 
Observatory millimeter-wave array.  They found a disk inclination of 21.5$^{+0.4}_{-0.3}\degr$ and PA of 58.6 $\pm$ 0.5$\degr$.
\citet{Pietu:2005aa} targeted AB Aur with the IRAM Plateau de Bure interferometer in CO and mm continuum intensity emission.  
They measured inclinations of $\approx$33$\degr$ to 42$\degr$ and PAs of --27$\degr$ to --31$\degr$,
but they note that these values may been impacted by unresolved structure and asymmetric emission.   Pietu et al. conclude that
an inclination of $\approx$23$\degr$ is most likely as it reconciles the measured velocities and the expected mass of AB Aur.
\citet{Tang:2012ih} confirm an inconsistency of the inclination measured on different spatial scales---from $\approx$42$\degr$ at large
scales to $\approx$23$\degr$ at the location of the dust ring at $\approx$100--200~AU---which could be explained by 
a warp in the disk.

We adopt a disk inclination of 23 $\pm$ 2$\degr$ and PA of the disk rotation axis of --36 $\pm$ 5$\degr$ based largely on 
observations from \citet{RiviereMarichalar:2020aa} and \citet{Tang:2017gr}, which probed spatial scales similar to 
AB Aur b.  Here the uncertainties have been estimated to encapsulate the most precise measurements.
The rotation axis PA implies a PA for the longitude of ascending node, $\Omega$, of 54 $\pm$ 5 $\degr$
assuming AB Aur b is nested in the 
disk\footnote{This assumption of coplanarity with the disk is reinforced by AB Aur b's resolved nature, possibly associated with a circumplanetary disk or envelope, and connections to larger scale features in the disk (e.g., \citealt{Currie:2022dd}; \citealt{Zhou:2022fa}).}.
The northeastern side of the disk is blueshifted and the southwestern side is redshifted (e.g.,
\citealt{Corder:2005aa}; \citealt{Tang:2012ih}; \citealt{Tang:2017gr}; \citealt{Speedie:2024aa}); combined with the counterclockwise orbital
motion of AB Aur b from \citet{Currie:2022dd}, this implies the brighter southeast portion of the disk is 
on the near side and the northwest portion of the disk is on on the far side.  AB Aur b therefore appears to be on the near side
of the disk.  As noted by \citet{Fukagawa:2004kp}, this geometry is reinforced by the larger-scale spiral structure of the disk, which would be trailing in this
configuration.

Finally, the orbital distance of AB Aur b, $r$, can be determined from the projected separation $\rho$
and its orbital phase using the Thiele-Innes elements.  Assuming a circular orbit, and converting the argument of periastron
and true anomaly ($\omega$ + $\nu$(t)) to measured angles of the disk and companion ($\Delta$PA), the de-projected orbital semi-major axis is

\begin{equation}{\label{eqn:physicalr}}
r = \frac{\rho }{\varpi \sqrt{\cos^2(\Delta \mathrm{PA}) + \cos^2(i) \sin^2(\Delta \mathrm{PA})} } \ \mathrm{AU},
\end{equation}

\noindent where $\rho$ is the projected separation in units of $''$, $\varpi$ is the system parallax in $''$;
$\Delta$PA (equal to $\lvert \mathrm{PA} - \Omega \rvert\ $) is the difference between the position angle of the companion (measured East of North), PA,  and the longitude of ascending node, $\Omega$; and
$i$ is the orbital (and, equivalently, the disk) inclination.  
Here $r$ is expressed in units of AU.

\begin{figure}
  \vskip -0.7 in
  \hskip -0.2 in
  \resizebox{3.8in}{!}{\includegraphics{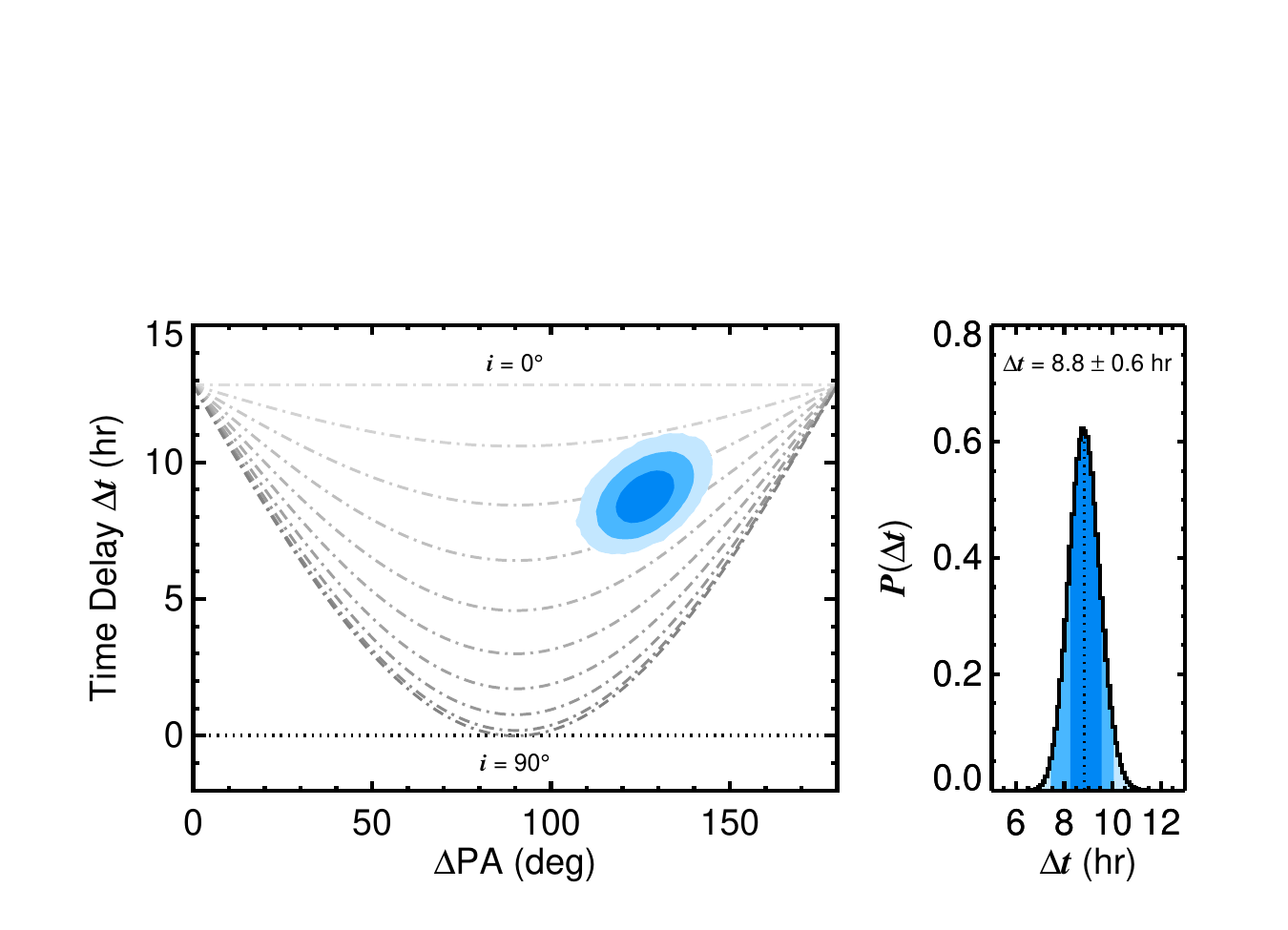}}
  \vskip -0.1 in
  \caption{Predicted delay time, $\Delta t$, for light echoes from AB Aur b.   Inclinations from 0$\degr$ to 90$\degr$ are shown in steps of 10$\degr$.  Left: The  
  protoplanetary disk inclination (23 $\pm$ 2$\degr$), orientation of the disk's rotation axis (--36 $\pm$ 5$\degr$), 
  and position of AB Aur b in its orbit 
  correspond to a constraint in the expected delay time.  These are represented here with 1-, 2-, and 3-$\sigma$ 
  contours in joint constraints of $\Delta t$ as a function of $\Delta$PA, the absolute difference between the 
  companion PA and the longitude of ascending node.   Right: Marginalized distribution delay times for AB Aur b. 
  Our best estimate is $\Delta t$ = 8.8 $\pm$ 0.6~hr.  \label{fig:abaurdt} } 
\end{figure}

Using the parallax to AB Aur from Gaia DR3 ($\varpi$ = 6.4127 $\pm$ 0.0372~mas; \citealt{GaiaCollaboration:2016cu}; \citealt{GaiaCollaboration:2022aa}),
the mean projected separation of 574 $\pm$ 15 mas from \citet{Zhou:2023di}, 
the companion PA of 181$\fdg$2 $\pm$ 2$\fdg$1 from \citet{Zhou:2023di}, 
and $\Omega$ and $i$ from our adopted values for the disk,
we find a true separation of $r$ = 94 $\pm$ 5~AU for AB Aur b.
This corresponds to a characteristic light travel time of $r/c$ = 13.1 $\pm$ 0.75 hr to reach AB Aur b,
and a \emph{delay} time of $\Delta t$ = 8.8 $\pm$ 0.6 hr based on Monte Carlo
realizations of $\Delta$PA and $i$ (Figure~\ref{fig:abaurdt}).

Here we have only considered the disk midplane where a planet would reside.  If the disk is flared, this will have a modest influence on the predicted 
arrival time of scattered light off of small grains in the disk atmosphere.  For AB Aur b, the impact would effectively be a higher inclination angle and a 
somewhat smaller delay time compared to the geometrically flat disk we have assumed.  See \citet{Gaidos:1994aa} for details of light echoes for a flared disk.


\begin{deluxetable*}{ccccccccc}
\renewcommand\arraystretch{0.9}
\tabletypesize{\small}
\setlength{ \tabcolsep } {.1cm}
\tablewidth{0pt}
\tablecolumns{9}
\tablecaption{HST WFC3/UVIS Imaging of AB Aur in F656N\label{tab:obs}}
\tablehead{
    \colhead{Orbit}  &  \colhead{Obs. Epoch}  &  \colhead{UT Date}  & \colhead{Mean Julian Date\tablenotemark{a}}       & \colhead{$\Delta$t\tablenotemark{b}}      & \colhead{Exp. Time}  &  \colhead{$N_\mathrm{Dith.}$ $\times$ $N_\mathrm{Exp.}$}   & \colhead{HST Roll Angle\tablenotemark{c}}   & \colhead{Notes\tablenotemark{d}} \\
          \colhead{Epoch ID}  &  \colhead{(UT)}       &  \colhead{(Y-M-D)}  & \colhead{(JD - 2457000)} &  \colhead{(hr)} &  \colhead{(s)}            &  & \colhead{($\degr$)}   &
        }
\startdata
E1                    &     2023.0190  &   2023-01-07   &   2952.43701    & 0.0   &  2.7 &  4 $\times$ 11 & 80.0  &  AB Aur A \\
E1                    &     2023.0203  &   2023-01-08   &   2952.89964    & 11.10 &  2.7 &  4 $\times$ 11 & 112.0 & AB Aur b  \\
E1                    &     2023.0205  &   2023-01-08   &   2952.96572    & 12.69 &  2.7 &  4 $\times$ 11 &  80.0  & AB Aur b \\
E1                    &     2023.0206  &   2023-01-08   &   2953.03181    & 14.28 &  2.7 &  4 $\times$ 11 &  112.0 & AB Aur b \\
E2                    &     2023.1040  &   2023-02-07   &   2983.44316    & 0.0   &  2.7 &  4 $\times$ 11 &  76.0 &  AB Aur A  \\
E2                    &     2023.1052  &   2023-02-08   &   2983.90499    & 11.08 &  2.7 &  4 $\times$ 11 &  103.5 & AB Aur b \\
E2                    &     2023.1054  &   2023-02-08   &   2983.97107    & 12.67 &  2.7 &  4 $\times$ 11 &  76.0  & AB Aur b  \\
E2                    &     2023.1056  &   2023-02-08   &   2984.03715    & 14.26 &  2.7 &  4 $\times$ 11 &  103.5 & AB Aur b \\
E3                    &     2023.1639  &   2023-03-01   &   3005.31240    & 0.0   &  2.7 &  4 $\times$ 11 &  70.0  &  AB Aur A \\
E3                    &     2023.1651  &   2023-03-02   &   3005.77483    & 11.10 &  2.7 &  4 $\times$ 11 &  95.0 & AB Aur b \\
E3                    &     2023.1653  &   2023-03-02   &   3005.84089    & 12.68 &  2.7 &  4 $\times$ 11 &  70.0 & AB Aur b \\
E3                    &     2023.1655  &   2023-03-02   &   3005.90695    & 14.27 &  2.7 &  4 $\times$ 11 &  95.0 & AB Aur b \\
E4                    &     2024.1406  &   2024-02-21   &   3361.97654    & 0.0   &  2.7 &  4 $\times$ 11 &  78.2  &  AB Aur A \\
E4                    &     2024.1419  &   2024-02-21   &   3362.43782    & 11.07 &  2.7 &  4 $\times$ 11 &  103.2 & AB Aur b \\
E4                    &     2024.1421  &   2024-02-22   &   3362.50372    & 12.65 &  2.7 &  4 $\times$ 11 &  78.2 & AB Aur b \\
E4                    &     2024.1423  &   2024-02-22   &   3362.56961    & 14.23 &  2.7 &  4 $\times$ 11 &  103.2  & AB Aur b \\
E5\tablenotemark{e}   &     2024.1749  &   2024-03-04   &   3374.49667    & 0.0   &  2.7 &  4 $\times$ 11 &  63.2  &  AB Aur A \\
E5                    &     2024.1763  &   2024-03-05   &   3375.02381    & 12.65 &  2.7 &  4 $\times$ 11 &  88.2 & AB Aur b \\
E5                    &     2024.1765  &   2024-03-05   &   3375.08969    & 14.23 &  2.7 &  4 $\times$ 11 &  63.2 & AB Aur b \\
E5                    &     2024.1767  &   2024-03-05   &   3375.15558    & 15.81 &  2.7 &  4 $\times$ 11 &  88.2 & AB Aur b \\
\enddata
\tablenotetext{a}{Average JD of the start of the first science observation and the end of the last observation for each orbit.}
\tablenotetext{b}{Elapsed time since first orbit in each four-orbit sequence.}
\tablenotetext{c}{Position angles are in HST's V3 coordinate system.}
\tablenotetext{d}{Notes about the intended science goals of each orbit.  Although all orbits have the same setup, the first orbit in each epoch
is designed to sample the host star flux, then after a gap in time, Orbits 2-4 are taken back-to-back to recover the candidate protoplanet AB Aur b.}
\tablenotetext{e}{For Orbit 1 of epoch E5, the Fine Guidance Sensors did not acquire the guide stars until after the observations were complete so guiding was carried out with GYRO pointing control.
However, this did not significantly impact the quality of the observations nor the results from this epoch.}
\end{deluxetable*}

\begin{deluxetable*}{lccccccccc}
\renewcommand\arraystretch{0.9}
\tabletypesize{\footnotesize}
\setlength{ \tabcolsep } {.1cm}
\tablewidth{0pt}
\tablecolumns{10}
\tablecaption{$F656N$ Photometry and Astrometry of AB Aur b\label{tab:photometry}}
\tablehead{
    \colhead{Epoch}  &  \colhead{Epoch}  &  \colhead{Host Count Rate\tablenotemark{b}}  & \colhead{Host $f_{\nu}$\tablenotemark{b}}    & \colhead{AB Aur b Count Rate\tablenotemark{c}}  &  \colhead{AB Aur b $f_{\nu}$\tablenotemark{c}}   & \colhead{P.A.}   & \colhead{$\rho$}  & \colhead{$\sigma_x$\tablenotemark{d}}  & \colhead{$\sigma_y$\tablenotemark{d}} \\
          \colhead{ID}  &  \colhead{(UT)}       &  \colhead{(e$^{-}$ s$^{-1}$)}  & \colhead{(mJy)} & \colhead{(e$^{-}$ s$^{-1}$)}    &  \colhead{(mJy)}     & \colhead{($\degr$)}   &   \colhead{($''$)}  &   \colhead{($''$)}  &   \colhead{($''$)}
        }
\startdata
       E0\tablenotemark{a} &  2022.17 &      440829 $\pm$ 6900   &        10540 $\pm$ 170   &            96 $\pm$ 11    &        2.29 $\pm$ 0.26    &        182.3 $\pm$ 1.1    &      0.595 $\pm$ 0.009    &            0.086    &            0.062    \\
       E1   &   2023.020              &       357720 $\pm$ 229   &           8554 $\pm$ 5   &            180 $\pm$ 9    &        4.30 $\pm$ 0.22    &          182.3 $\pm$ 1.0    &      0.594 $\pm$ 0.007    &             0.11    &            0.071    \\
       E2   &   2023.105              &       363580 $\pm$ 610   &          8694 $\pm$ 15   &             65 $\pm$ 9    &        1.55 $\pm$ 0.21    &        178.3 $\pm$ 1.2    &      0.582 $\pm$ 0.008    &            0.092    &            0.054    \\
       E3   &   2023.165              &       331200 $\pm$ 730   &          7920 $\pm$ 17   &             64 $\pm$ 6    &        1.52 $\pm$ 0.15    &        180.0 $\pm$ 1.0    &      0.583 $\pm$ 0.008    &            0.085    &            0.058    \\
       E4   &   2024.142              &       373650 $\pm$ 380   &           8935 $\pm$ 9   &           272 $\pm$ 18    &        6.50 $\pm$ 0.43    &          181.7 $\pm$ 1.0    &      0.591 $\pm$ 0.008    &             0.11    &            0.082    \\
       E5  &   2024.176               &       324590 $\pm$ 550   &          7762 $\pm$ 13   &           148 $\pm$ 10    &        3.53 $\pm$ 0.23    &        184.4 $\pm$ 1.0    &      0.594 $\pm$ 0.007    &            0.094    &            0.057    \\
\enddata
\tablenotetext{a}{``E0'' is a new reduction of HST WFC3/UVIS H$\alpha$ observations from \cite{Zhou:2022fa}.  These data did not include a delayed sampling of the host star, AB Aur, and the candidate companion AB Aur b.}
\tablenotetext{b}{Measured from Orbit 1 of each epoch, with the exception of E0.}
\tablenotetext{c}{Measured from Orbits 2--4 of each epoch, with the exception of E0.}
\tablenotetext{d}{Standard deviations of the best-fitting 2-D Gaussian model used in forward modeling.  These represent widths of the input model, not the deconvolved implied angular size of AB Aur b.}
\end{deluxetable*}

\section{Observations}{\label{sec:obs}}

\subsection{TESS Light Curves of AB Aur}{\label{sec:tess}}

AB Aur was observed by TESS in Sectors 19, 43, 44, 59, 71, and 86. We downloaded the light curves of each sector from the Mikulski Archive for Space Telescopes (MAST) data archive. For Sector 19, only full-frame image (FFI) light curves are available (\citealt{Caldwell:2020aa}). For Sectors 43, 44, 59, 71, and 86 we used the \texttt{lightkurve} (\citealt{LightkurveCollaboration:2018aa}) software package to obtain the 2-minute cadence TESS light curve (\citealt{Smith:2012aa}; \citealt{Stumpe:2012aa}; \citealt{Stumpe:2014aa}; \citealt{Jenkins:2016aa}).\footnote{TESS FFI and 2-minute cadence light curves are available at MAST: \citet{https://doi.org/10.17909/t9-wpz1-8s54} 
and \citet{https://doi.org/10.17909/t9-nmc8-f686}
respectively.} All photometry listed as \texttt{NaN} are first removed. Then, light curves from all sectors are stitched together and median normalized by dividing the flux and flux errors by the median flux of the combined light curve.

\subsection{Hubble Space Telescope WFC3/UVIS Imaging}

Deep H$\alpha$ imaging of AB Aur was carried out with HST (GO 17168; PI: Bowler) across five epochs spanning 20 orbits from January 2023 to March 2024
(UT 2023 January 7--8, 2023 February 7--8, 2023 March 1--2, 2024 February 21--22, March 4--5) 
using WFC3's ultraviolet and visible light (UVIS) channel and the \emph{F656N} narrow-band filter 
($\lambda_\mathrm{c}$ = 6561.4~\AA; $\Delta \lambda$ = 17.6~\AA)\footnote{The data is available at MAST: \href{http://dx.doi.org/10.17909/cve2-6v41}{http://dx.doi.org/10.17909/cve2-6v41}.}
Each epoch consists of four orbits; the first orbit is designed to measure the H$\alpha$ flux emitted 
from the host star followed by a second visit of three contiguous orbits at a later time to sample AB Aur b.
These three orbits during the second visit are required to reach a sufficient total exposure time and depth to
recover AB Aur b in a similar fashion as in \citet{Zhou:2022fa}.
Observations for other HST programs are taken in between these visits.

The spacing of the first and second visits is intended to capture the time delay
associated with light traveling from the host star to AB Aur b and scattering to an observer, which depends on the physical
separation of the companion and its orbital geometry as described in Section \ref{sec:abaurgeometry}.
HST's orbital period is about 96~min, with typical blocks of about 50~min available for guide-star acquisition and science observations.
This means the ``effective time resolution'' of each 3-orbit set of observations focused on AB Aur b from the beginning of 
Orbit 2 to the end of Orbit 4 for each epoch is about 242 min (96 min $\times$ 2 + 50 min),
or roughly 4 hours, with the midpoint occurring about 2 hours into the sequence.  
The expected time delay for the light echo from AB Aur b to an observer is about 9 hours (Figure~\ref{fig:abaurdt}).
Given the uncertainties in the geometry of the system, the start of the observations are within 2--3~$\sigma$ of the predicted timing of the light echo for each of the five epochs.\footnote{If the inclination
is larger than expected, this will reduce the time delay and increase the tension between the predicted time and the observation start time.
For instance, a 42$\degr$ inclination gives a predicted delay time of 6.0~hours.}
The gap in time between the first (Orbit 1) and second (Orbits 2--4) visits from this program was about 10 hours (for epochs E1--E4) and 12 hours for one epoch (E5), corresponding to a wait time of 6--7 orbits.  This is
slightly longer than the expected time delay by about 1--3 hours from the end of Orbit 1 to the start of Orbit 2,
or 3.5--5.5 hours from the midpoint of Orbit 1 to the midpoint of Orbit 3.
Nevertheless, this modest ``overshooting'' is an improvement over a scenario with no gap in the four orbit scheduling of each epoch, which
would have resulted in an underestimate of 6.5 hours, or in which 
the host star flux was directly obtained from the 3 orbits dedicated for the companion, which would have underestimated the time delay by 9 hours.

Special scheduling constraints are imposed to avoid diffraction spikes from AB Aur falling at the position 
of the companion, located at a separation of 0$\farcs$6 from the host star.
Each 4-orbit group is taken with relative roll angle requirements between 25--35$\degr$
in an A-BAB pattern (representing Orbit 1 followed by Orbits 2--4 at a later time) 
to facilitate PSF subtraction through angular differential imaging (ADI; \citealt{Liu:2004kk}; \citealt{Marois:2006df}).
Here ``A" refers to one roll angle orientation and ``B'' refers to the second roll angle.
Actual roll angles range from 25--32$\degr$ among the five epochs.
Images are taken with the UVIS2-C512C-SUB 512$\times$512 pix subarray, resulting in a field of view of 
20$\farcs$5$\times$20$\farcs$5 for the WFC3/UVIS plate scale of 40~mas pix$^{-1}$.
The same setup is used for all four orbits of each epoch.


\begin{figure*}
  \vskip -0.6 in
  \hskip -0.7 in
  \resizebox{8.5in}{!}{\includegraphics{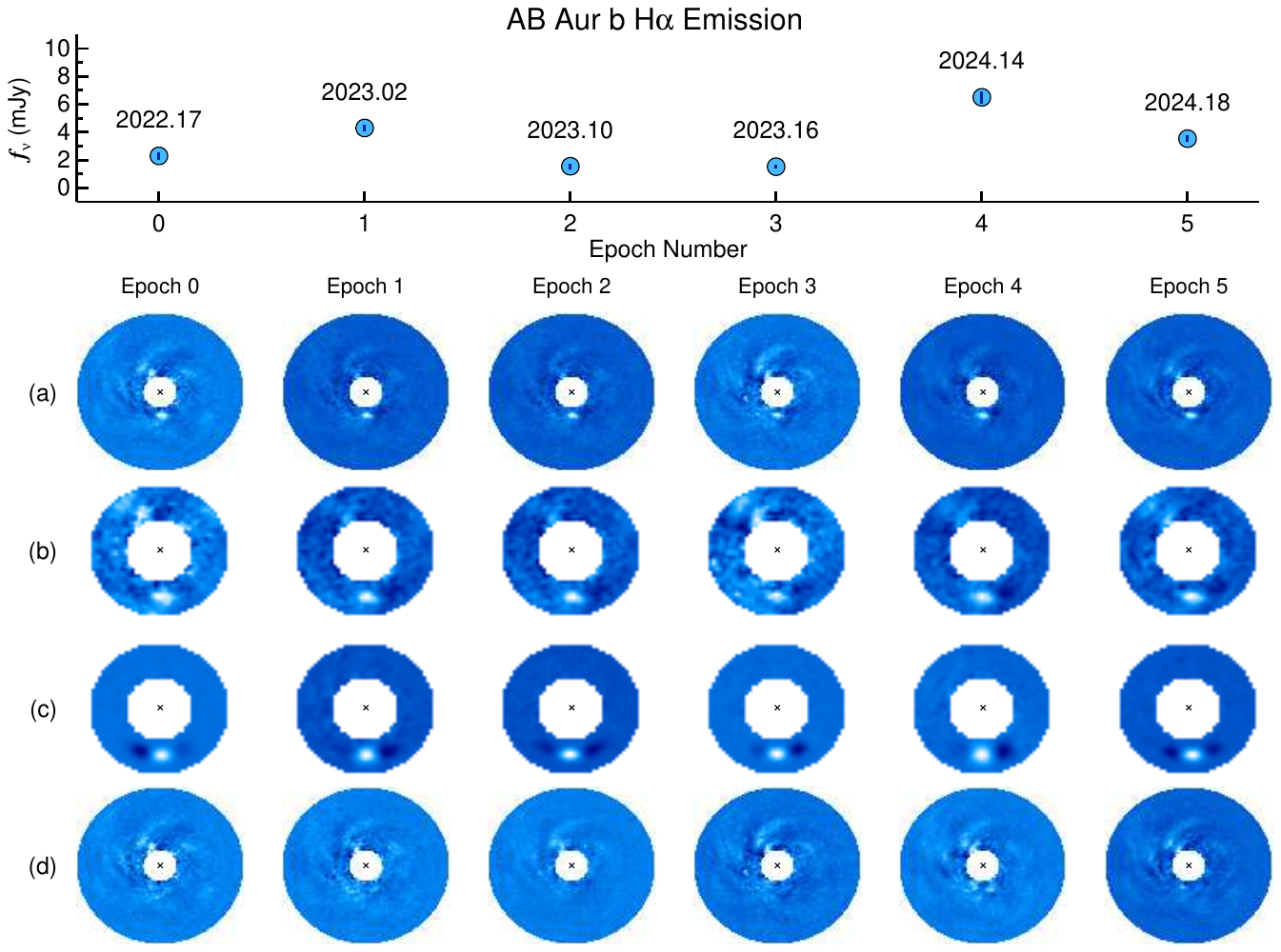}}
  \vskip -0.65 in
  \caption{HST/WFC3-UVIS H$\alpha$ (\emph{F656N} filter) observations of the candidate protoplanet AB Aur b.  
  Epochs E1--E5 originate from our accretion light echo
  experiment, which spans a 14-month time baseline in 2023--2024, 
  while epoch E0 (originally presented in \citealt{Zhou:2022fa}) was acquired in 2022 and has been reprocessed here in the same way as the new epochs.  
  \emph{Upper panel:} H$\alpha$ flux density in mJy as a function of epoch number.  AB Aur b shows strong variability 
  at the 14.7$\sigma$ level, with brightness increasing by 330\% from E3 to E4; see Section~\ref{sec:abaurbvar} for details.
  \emph{Panel (a):} Primary-subtracted images of AB Aur for epochs E0--E5.  AB Aur b as well as fainter disk features are clearly visible in each frame.
  Orbits 2, 3, and 4 are used for frames E1--E5, while E0 represents all five orbits from that program (GO-16651).
  \emph{Panel (b):} Zoom-in on AB Aur b, located 0$\farcs$6 south of the host star.  
   The dark lobes on either side of the source are artifacts caused by self-subtraction of AB Aur b 
  using frames acquired at  different roll angles.
  \emph{Panel (c):} Best-fit 2D Gaussian model of AB Aur b from the forward-modeling procedure described in Section~\ref{sec:astrphot}.  
  AB Aur b is resolved, with greater extended emission in the azimuthal direction.
  \emph{Panel (d):} Residuals after subtracting the best-fit model in Panel (c). 
  North is up and East is to the left in all images.  For scale, the inner radius is 0$\farcs$4 (62 AU) and the outer annulus in Panels (a) and (d) is 2$\farcs$0 (312 AU).
   \label{fig:abaurepochs} } 
\end{figure*}

A WFC3-UVIS-DITHER-BOX 4-point dither pattern is employed with 0$\farcs$02 spacing between positions
for cosmic ray rejection and to mitigate bad pixels.
For each dither pointing, 11 images in $F656N$  are acquired, each with an exposure time of 2.7~s,
resulting in 880 frames altogether and a total on-source integration time of 2376~s (40 min).
However, as the first orbit of each epoch is intended to sample the host star's H$\alpha$ emission,
only 132 frames are used for the reduced image of AB Aur b, 
amounting to an exposure time of 356~s (5.9~min) at each epoch.
A minimum of 10 days between epochs is imposed to ensure AB Aur's H$\alpha$ variability
is sampled on timescales that are uncorrelated with the previous epoch.
The first three epochs were taken within a three month period in early 2023, and the 
last two epochs were obtained in early 2024.  The corresponding epoch-to-epoch baselines range from  12 days to nearly one year.
Details of the observations can be found in Table~\ref{tab:obs}.

In addition to new observations taken as part of this accretion light
echo program, we also uniformly re-reduce previously obtained H$\alpha$ observations of AB Aur taken with WFC3/UVIS  
in F656N (GO 16651; PI: Zhou). 
These data were presented in \citet{Zhou:2022fa} and span five orbits across two visits separated by nearly two months: 
three orbits on UT 2022 February 7 
and two orbits on UT 2022 March 28.  
Although this dataset did not include a light travel
time delay between the host and candidate companion as in this survey, 
the observations were otherwise carried out in a similar fashion and offer an opportunity to analyze longer-term variability
about 10 months prior to the first epoch from this program.
We refer to this epoch from 2022.17 as Epoch 0 (``E0'') and the five epochs from this survey spanning 2023.02 to 2024.176 as Epochs 1--5 (``E1'' to ``E5'').

\subsection{WFC3 Data Reduction}{\label{sec:wfc3dr}}

The data processing largely follows procedures described in \citet{Zhou:2022fa} and \citet{Zhou:2023di}.
The reduction starts with the \texttt{flc} FITS files downloaded from the MAST archive. 
These files are processed by the \texttt{calwf3} pipeline, which includes calibrations of bias, flat field, and 
charge transfer efficiency corrections. Cosmic rays and bad pixels are identified from the calibrated images and replaced by bilinear interpolation of neighboring pixels.
A small number of bad pixels persist throughout each epoch and are flagged using Data Quality (DQ) values 1024 (charge trap sink pixels) and 4096 (cosmic rays detected by \texttt{AstroDrizzle}), 
with remaining spurious pixels identified by eye in epoch-averaged images. We find that this approach is more robust than using DQ flags alone. There are no bad pixels apparent near the companion.
Images are registered and assembled into data cubes by aligning the centroid positions of AB Aur to a common origin
through sub-pixel image resampling.
The native plate scale of 40~mas pixel$^{-1}$ is preserved throughout the image processing and analysis.

Photometry of the host star is measured for every image taken within the first of four orbits of each epoch using aperture photometry
with a radius of 10 pixels (0$\farcs$4).  A correction is then applied for UVIS2 encircled energy loss with the F656N 
filter\footnote{\url{https://www.stsci.edu/hst/instrumentation/wfc3/data-analysis/photometric-calibration/uvis-encircled-energy}}.
Count rates are converted to flux densities using the inverse sensitivity conversion factors contained in the 
\texttt{PHOTFLAM} FITS header keyword.  Average count rates and flux densities 
of AB Aur are presented in Table~\ref{tab:photometry}.  Intra-orbit and inter-orbit light curves are further discussed
in Appendix~\ref{sec:appa}.

PSF subtraction is carried out 
using the 
Karhunen-Lo\`{e}ve Image Projection (KLIP; \citealt{Soummer:2012ig}) algorithm implemented with the \texttt{pyKLIP} 
package (\citealt{Wang:2015zz}).  Following \citet{Zhou:2023di}, each science frame makes use of reference images
of all other observations with a minimum roll angle difference of 20$\degr$.  The number of KL modes is set to 30 for all reductions.
KLIP is applied in seven annuli spanning 0$\farcs$1 to 2$\farcs$0.  After PSF subtraction, each frame is north aligned, then
all observations from Orbits 2--4 of each epoch are coadded into a single processed image.

\subsection{Astrometry and Photometry}{\label{sec:astrphot}}

AB Aur b is clearly recovered in the primary-subtracted frames for all five epochs (Figure~\ref{fig:abaurepochs}).  
The morphology of the source appears  similar to its appearance in primary-subtracted 
HST images from \citet{Zhou:2023di}
spanning the UV through optical wavelengths.
AB Aur b is the brightest source in the processed frames and is clearly extended 
in the azimuthal (East-West) direction.

To account for self-subtraction from the ADI processing,
photometry and astrometry are measured in \texttt{pyKLIP} by forward modeling a 2D Gaussian assuming independence
of the $x$ and $y$ spatial directions (that is, no covariance is included).
The Gaussian standard deviations ($\sigma_x$ and $\sigma_y$) are determined from direct fits to the primary-subtracted image,
Then, three-parameter posteriors are sampled with Markov-chain Monte Carlo: the central position of the source and 
the amplitude.
Uniform priors are adopted for each parameter, and the median and standard deviation of the posterior chains
are adopted for the reported central values and corresponding uncertainties.  
Results are listed in Table~\ref{tab:photometry}, and the best-fitting models and residuals of the 
fits for each epoch are shown in Figure~\ref{fig:abaurepochs}.

The inferred position angle ranges from 178.3$\degr$ (E2) to 184.4$\degr$ (E5), and the 
separation is about 0$\farcs$6 in all epochs.  These are in good agreement with previous measurements in
 \citet{Currie:2022dd}, \citet{Zhou:2022fa}, and \citet{Zhou:2023di}.
The dispersion in the azimuthal direction is consistently larger than that in the radial direction,
with an average value of $\sigma_x$ = 0$\farcs$096, which subtends an angle of 9$\degr$ at separation of 0$\farcs$6 from the host star,
and an average of $\sigma_y$ = 0$\farcs$064 in the radial direction.
The resulting aspect ratio is $\sigma_x / \sigma_y$ = 1.5.
For comparison, the  HST diffraction limit at H$\alpha$ is 29~mas (FWHM = 68~mas); AB Aur b is therefore spatially 
resolved in both directions.  
The deconvolved size of AB Aur b is 91~mas and 57~mas in the azimuthal and radial directions, respectively,
corresponding to a physical size of 14~AU $\times$ 9~AU at the distance of AB Aur.
This is larger than the upper limits of $\sim$1--5~AU for the other spatially unresolved circumplanetary disks surrounding PDS 70 c and SR 12 c (\citealt{Benisty:2021ft}; \citealt{Wu:2022aa}).
Because the source is spatially resolved, the positional uncertainty of AB Aur b is likely to be underestimated 
when considering only the spread in the posterior distributions of the $x$ and $y$ position from the forward modeling process.  
We therefore include an additional spread of 10\% of the average of $\sigma_x$ and $\sigma_y$ 
in quadrature with the inferred positional
uncertainty of AB Aur b as a conservative estimate of potential systematic errors in astrometry.
This 10\% value represents the approximate scale of the difference in dispersion estimates from the individual reduced frames.
These inflated errors are included in the astrometric measurements listed in Table~\ref{tab:photometry}.

Note that the recovered astrometry and photometry of AB Aur b as part of our newly reduced E0 epoch
differs slightly from---but is formally consistent with---measurements of the same dataset from  \citet{Zhou:2022fa}.
Here we find a separation of 595 $\pm$ 9~mas and a P.A. of 182$\fdg$3 $\pm$ 1$\fdg$1,
which is in good agreement with the separation of 600 $\pm$ 22~mas and P.A. of 182$\fdg$5 $\pm$ 1$\fdg$4 from \citet{Zhou:2022fa}.
On the other hand, we recover slightly higher value for the brightness of AB Aur b; we measure
a flux density of 2.29 $\pm$ 0.26~mJy, compared with 1.5 $\pm$ 0.4~mJy from \citet{Zhou:2022fa}.
Although our value is 53\% larger, given the uncertainties, the two measurements are nevertheless
consistent at the 1.7$\sigma$ level.
The difference can be attributed to the point-source model used in \citet{Zhou:2022fa} compared with the more accurate
extended source model that we adopt here.

\section{Results}{\label{sec:results}}

\subsection{Strong H$\alpha$ Variability from AB Aur b}{\label{sec:abaurbvar}}

One simple approach to assess variability of brightness measurements 
is to compare the characteristic uncertainty of individual measurements, $\bar{\sigma_{f_\nu}}$, to the 
sample standard deviation, $s = \sqrt{1/(N-1) \cdot \sum_{i=1}^N{(f_{\nu,i}-\bar{f_{\nu}})^2}}$,
where $N$ is the number of data points, $f_{\nu,i}$ is the $i$th brightness measurement, and $\bar{f_{\nu}}$ is the sample mean.  
A ratio $\bar{\sigma_{f_\nu}}$/$s$ near unity would be expected for data without significant variability,
whereas a ratio $\ll$1 points to real changes.  
For the five epochs spanning 14 months sampled with this HST program, $\bar{\sigma_{f_\nu}}$/$s$ = 0.11.  Including
E0, this ratio is 0.12\footnote{Note that this analysis does not account for slight reddening to AB Aur, which was estimated by \citet{McJunkin:2014aa} to be $A_V$=0.51~mag.  Extinction to AB Aur b could be significantly greater if it is indeed an embedded protoplanet.}.
AB Aur b shows clear epoch-to-epoch brightness changes that are a factor of 8--10 higher than would be expected from
measurement scatter alone.

We can further quantify the significance of variability by computing the $\chi^2$ value of a constant model assuming no variability 
as the null hypothesis
and comparing this to the expected range of $\chi^2$ values for the appropriate degrees of freedom, $\nu$.
Considering all six epochs, we find a $\chi^2$ value of 235 for $\nu$ = $N$--1 = 5, 
where $\chi^2$ = $\sum_{i=1}^{N} (f_{\nu,i}-c_{\nu})^2 / \sigma_{f_{\nu,i}}^2$.
Here $c_{\nu}$ is the optimally scaled constant model which minimizes $\chi^2$ and can be found
analytically for model values $m_{\nu,i}$: 
$c_{\nu} = \sum_{i=1}^{N} (f_{\nu,i}  m_{\nu,i} / \sigma_{f_{\nu,i}}^2) / \sum_{i=1}^{N} (m_{\nu,i}^2 / \sigma_{f_{\nu,i}}^2)$ (\citealt{Cushing:2008kb}).
When $m_{\nu,i}$ is an arbitrary constant, for instance 1, then the scale factor is 
$c_{\nu} = \sum_{i=1}^{N} (f_{\nu,i} / \sigma_{f_{\nu,i}}^2) / \sum_{i=1}^{N} (1 / \sigma_{f_{\nu,i}}^2)$, which is the weighted mean
of flux density measurements.  For AB Aur b, $c_{\nu}$ is 2.58 mJy.

The SNR in effective standard deviations of a Gaussian distribution is 
$n$=$\sqrt{2}$ erf$^{-1} (\int_{0}^{\chi^{2}} \chi^{2'}(x|\nu=5) dx)$,
where $\chi^{2'}(x|\nu=5)$ is the probability density function of the chi-square distribution for 5 degrees of freedom\footnote{For a Gaussian distribution, 
the probability that a value falls within $\pm$$n$ standard deviations from the mean is erf($n$/$\sqrt{2}$), where
erf is the error function defined as erf($z$) $\equiv$ 2/$\sqrt \pi $ $\int_0^z e^{-t^2} dt$. }.
For a $\chi^2$ value of 235, this constant-brightness model can be ruled out in favor of real variability at the 14.7$\sigma$ level.
Although we have no reason to believe our photometric uncertainties are underestimated, \emph{if} the 
errors were to be inflated by a factor of 2, the constant model would result in a $\chi^2$ value of 58.9 and 
would still be strongly disfavored at the 6.7$\sigma$ level.

The two faintest epochs are E2 and E3, which were taken within one month of each other from February to March 2023 and
have similar flux densities of 1.55 $\pm$ 0.21~mJy and 1.52 $\pm$ 0.15 mJy, respectively.
The brightest epoch is E4 with a flux density of 6.50 $\pm$ 0.43 mJy, taken in February 2024.
This represents an overall brightening by a factor of 4---specifically, a 330 $\pm$ 50\% increase from its faintest state---over 
a time baseline of about one year between the observations.
The re-reduction of the E0 dataset from 2022 (ten months prior to E1) yields a flux density that is overall comparable to the range that is seen from 2023--2024.

Variations over much shorter timescales are also observed across the coarse sampling of this program.
AB Aur b dimmed by 64\% from 4.30 $\pm$ 0.22~mJy to 1.55~$\pm$~0.21~mJy over a one month period
between epochs E1 and E2.  It similarly dimmed by 46\% from 6.50 $\pm$ 0.43~mJy to 3.53~$\pm$~0.23~mJy 
from E4 to E5 over 13 days.

\subsection{Modest H$\alpha$ Variability from AB Aur}

The host star AB Aur also shows clear signs of H$\alpha$ variability in the HST dataset.  Here we first 
consider variability within the first orbit for epochs E1--E5, and similarly for the first of the five orbits for E0.
Overall the brightness changes \emph{within} 
each $\approx$50-min epoch are small.  For E1--E5, AB Aur is variable at the sub-percent level
with an average standard deviation of 0.14\%.
This suggests that appreciable changes in mass accretion are occurring on longer timescales.  
E0 consists of 5 orbits across two visits over two months; the host star flux density in Table~\ref{tab:photometry}
is the mean over all five orbits and the quoted relative uncertainty is 1.6\%---an order of magnitude larger than for the other epochs,
likely a result of the time gap between Orbits 1--3 and Orbits 4--5 for that epoch.

On the other hand, AB Aur shows more substantial epoch-to-epoch variability.
The overall scatter is 11.4\% for E0--E5, and 6.1\% for E1--E5. 
AB Aur experiences a brightness change by a factor of 1.15, or 15\%, from minimum (E5) to maximum (E0).
This is similar to the expected level of H$\alpha$ variability based on previous spectroscopic monitoring campaigns 
(e.g., \citealt{Harrington:2007bq}; \citealt{Costigan:2014aa}), and is comparable to the 15\%
change measured by \citet{Zhou:2022fa} over the course of about two months.
It is also comparable to the continuum variability seen in the TESS light curves, which are taken before E0 (Sectors 19, 43, and 44),
between E0 and E1 (Sector 59), between E3 and E4 (Sector 71), and after E4 (Sector 86).
The uncertainty-to-scatter ratio, $\bar{\sigma_{f_\nu}}$/$s$, is 0.015 across all six epochs and 0.03 for E1--E5, a sign that 
the observed scatter exceeds measurement uncertainties by a factor of 30--70.
Following the same framework from Section~\ref{sec:abaurbvar} to assess the significance level of the observed variability,
we find a $\chi^2$ value of 6947 for a constant-flux model, which corresponds to a $\sim$80$\sigma$ detection of
variability for $\nu$=5 degrees of freedom.
This does not seem to be driven by E0, which is the most discrepant epoch; excluding E0 gives $\chi^2$=6800 ($\nu$=4)
with a comparable SNR.


\begin{figure}
  \vskip -0.5 in
  \hskip -0.6 in
  \resizebox{5in}{!}{\includegraphics{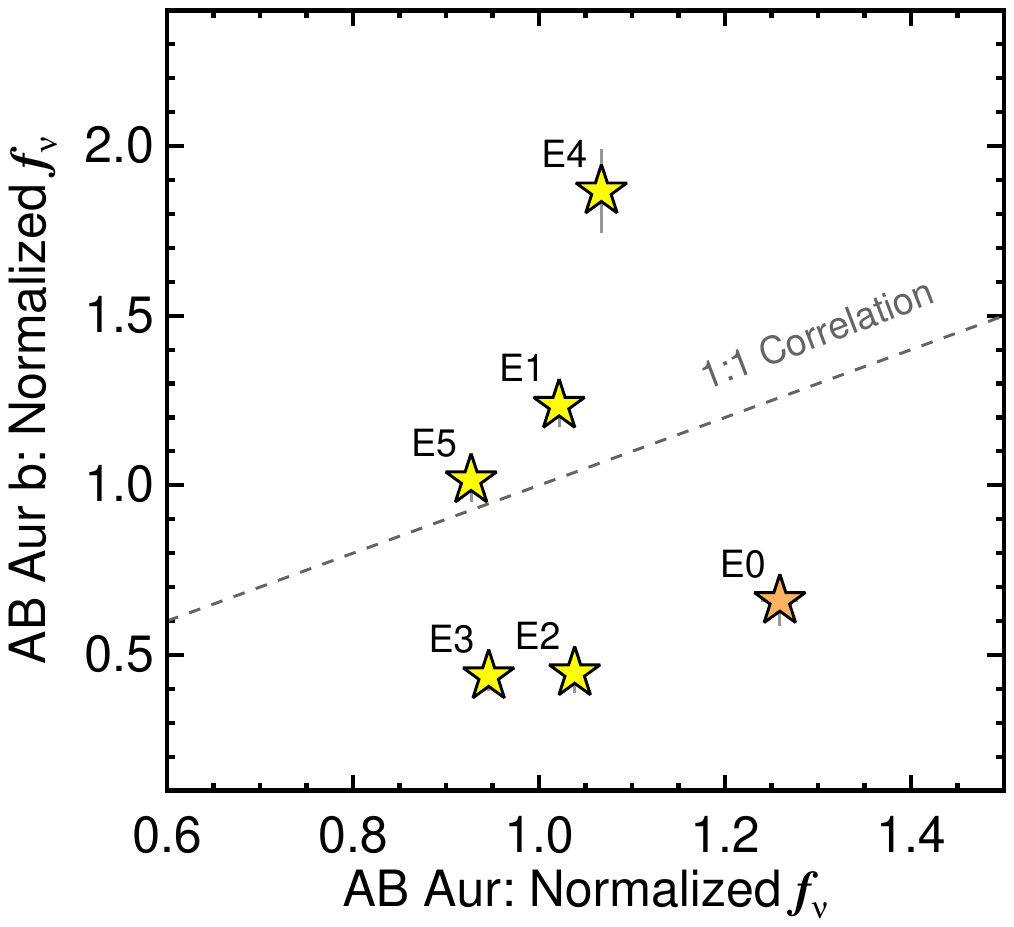}}
  \vskip -0.4 in
  \caption{Comparison between normalized HST H$\alpha$ photometry of AB Aur and 
  delayed brightness measurements of the candidate protoplanet AB Aur b 
  (yellow stars).
  If AB Aur b is an unobscured static disk feature, a positive linear correlation would be expected between AB Aur and AB Aur b.
  Instead, in this accretion 
  light echo experiment  we find that AB Aur varies by about 15\% while AB Aur b varies by 330\% (epochs E1--E5).  
  For comparison, our new reduction of the HST ``E0'' epoch from \citet{Zhou:2022fa} is displayed with an orange star.
  Note, however, that this dataset does not include the time delay between the host and candidate accreting planet.
  All measurements have been normalized to the mean values of the host star and companion from epochs E1--E5.
   \label{fig:onetoone} } 
\end{figure}

\subsection{AB Aur and AB Aur b have Uncorrelated Variability}

AB Aur b is clearly much more variable than the host star: AB Aur exhibited 15.1 $\pm$ 0.2\% brightness changes throughout this 
program while AB Aur b experienced changes at the 330 $\pm$ 50\% level---a factor of 22 $\pm$ 3 higher than the host.
This alone is a strong sign that emission from these two objects is unrelated.
Despite the inconsistent amplitudes, is this variability of the host star correlated with that of AB Aur b?  
Figure~\ref{fig:onetoone} compares the normalized  photometry of the host and companion 
for all five epochs of this accretion light echo experiment.  If AB Aur b is a static scattered-light disk feature, 
an approximately linear correlation would be expected because changes in host star accretion would be
directly and proportionately mimicked in the delayed images of AB Aur b.
This is not what is observed; epochs E1--E5 significantly depart from a 1:1 correlation by 
21\% for E1, 57\% for E2, 54\% for E3, 75\% for E4, and 9\% for E5.
Furthermore, there is no evident pattern in the variations; E2 and E3 are fainter than the 1:1 trend line, while E1, E4, and E5 are brighter.  
We conclude that the H$\alpha$ emission from AB Aur b cannot \emph{only} be unobstructed scattered light from the host star.

For comparison, we also include the location of E0 in Figure~\ref{fig:onetoone}, although we emphasize that those observations
were not taken as part of this accretion light echo experiment, and when comparing the host brightness to that of the
companion, cannot be evaluated in the same self-consistent manner.  Nevertheless, we note that 
for E0, the host star was appreciably brighter and the companion fainter than in epochs E1--E5 
and is another example of substantial departure from what might
be expected in the case that AB Aur b is a disk feature seen in scattered light.

The Pearson sample correlation coefficient between the host and companion for E1--E5 is $r$=0.49 with a relatively large
confidence interval of $\sigma_r$=0.59 using bootstrap resampling to estimate an uncertainty and $\sigma_r$=0.32 using jackknife resampling; both
approaches give comparably broad ranges that are consistent with no correlation.
However, the correlation coefficient should be treated with caution because the
number of measurements (five) is much smaller than is generally recommended for a robust 
interpretation of this metric (\citealt{Bujang:2016aa}).
Moreover, the large uncertainties are consistent with both strong positive correlation ($r$=1) and complete independence
of each parameter ($r$=0) within about 1$\sigma$.   


\begin{figure}
  \vskip -0.5 in
  \hskip -0.65 in
  \resizebox{5.2in}{!}{\includegraphics{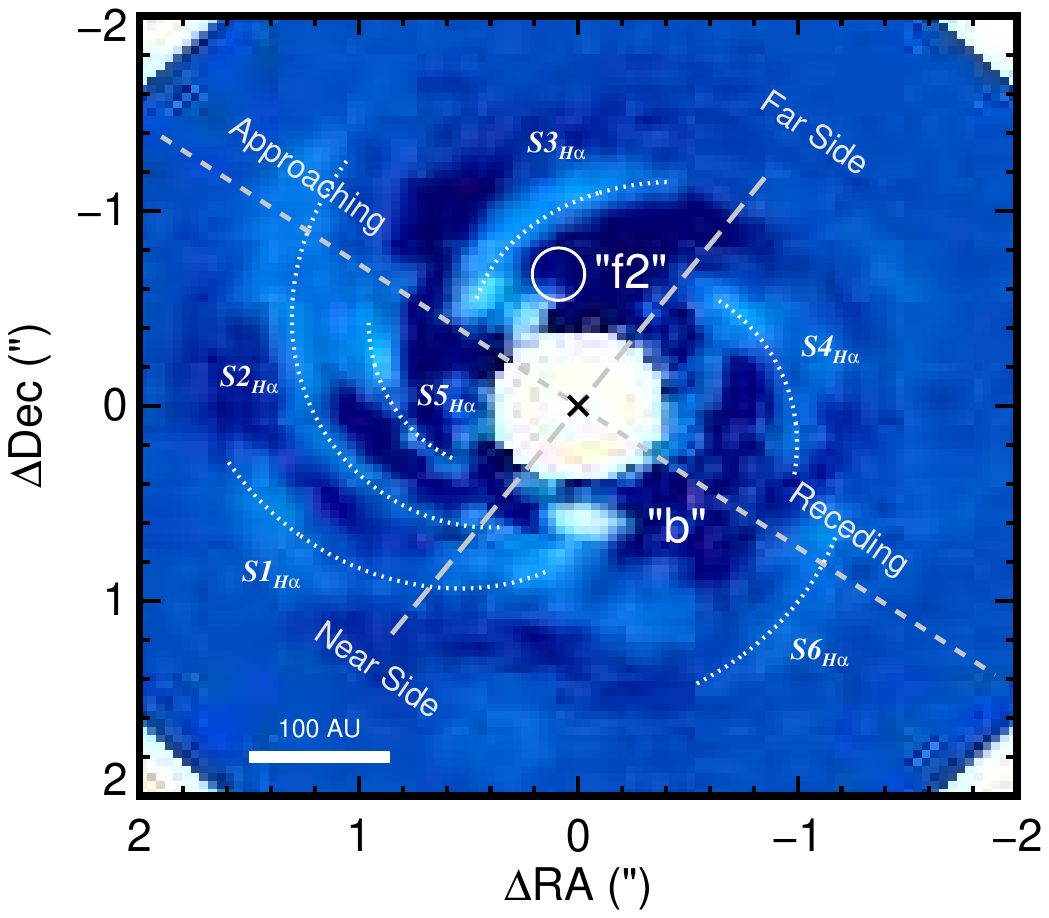}}
  \vskip -0.55 in
  \caption{Coadded \emph{F656N} image of AB Aur with HST.
  Observations represent the average of 6 epochs (E0--E5) of high-contrast imaging in H$\alpha$, amounting to a total integration time of 2.5 hr
  from 25 orbits over two years (2022--2024).  
  The short-dashed line is the line of nodes---the intersection of the sky and disk planes---and the long-dashed line is the projected rotation axis of the
  disk based on ALMA and NOEMA observations (\citealt{Tang:2017gr}; \citealt{RiviereMarichalar:2020aa}).  The disk is rotating counter-clockwise, with the north-east portion approaching and the south-east portion receding.
  AB Aur b (\citealt{Currie:2022dd}; \citealt{Zhou:2022fa}) is readily visible south of AB Aur (black $\times$), but the point source \emph{f2}
  north of AB Aur from \citealt{Boccaletti:2020ba} is not recovered.  Known H$\alpha$ spiral features ($S1_{H\alpha}$---$S4_{H\alpha}$) and new scattered-light 
  disk structures ($S5_{H\alpha}$ and $S6_{H\alpha}$) are labeled.  No new H$\alpha$ point sources are identified.
 \label{fig:coadded} } 
\end{figure}

\subsection{H$\alpha$ Line Flux, Accretion Luminosity, and Mass Accretion Rate}{\label{sec:mdot}}

Accretion luminosities and mass accretion rates can be derived from the observed H$\alpha$ emission of AB Aur and AB Aur b.
For each epoch, flux densities in $F656N$ from Table~\ref{tab:photometry} are corrected for
extinction assuming a reddening of $A_V$=0.51~mag to AB Aur (\citealt{McJunkin:2014aa}).
The  \citet{Fitzpatrick:1999dx} Galactic extinction curve and a ratio of total to selective extinction of $R_V$=3.1 are adopted.
For AB Aur, the contribution of the stellar continuum to the $F656N$ photometry is removed
by scaling a $T_\mathrm{eff}$=9800~K and log$g$=4.0~dex (\citealt{Soubiran:2016aa}) PHOENIX/Gaia model atmosphere (\citealt{Brott:2005vw}) 
to the $V$-band photometry of AB Aur, carrying out synthetic photometry with the HST $F656N$ transmission curve (which yields a monochromatic
flux density of 2.07 $\times$ 10$^{-12}$ erg$^{-1}$ s$^{-1}$ cm$^{-2}$ \AA$^{-1}$),
and subtracting this from the extinction-corrected values.  The resulting
H$\alpha$ line flux and corresponding H$\alpha$ line luminosity ($L_\mathrm{H \alpha}$) adopting a
Gaia DR3 distance of 155.9 $\pm$ 0.9~pc is reported for each epoch in Table~\ref{tab:acclum}.
Note that this treatment of accretion luminosity assumes that chromospheric H$\alpha$ emission is negligible in comparison 
at this early evolutionary phase of growth (e.g., \citealt{Manara:2017aa}).

Accretion luminosity ($L_\mathrm{acc}$) can be inferred using empirical trends between $L_\mathrm{H \alpha}$ and $L_\mathrm{acc}$.
Using the $L_\mathrm{H \alpha}$-$L_\mathrm{acc}$ relation from \citet{Mendigutia:2011aa}, which is calibrated for Herbig AeBe stars, yields
accretion luminosities between log($L_\mathrm{acc}$/$L_{\odot}$) = 1.1 and 1.2  for AB Aur, with a weighted mean of 
log($L_\mathrm{acc}$/$L_{\odot}$) = 1.12 $\pm$ 0.12  across the six HST epochs.
The mass accretion rate can then be found from $L_\mathrm{acc}$ = $G M_*$$\dot{M}$/$R_*$.
Assuming a stellar mass of $M_*$ = 2.4~$\pm$~0.2~$M_{\odot}$ (\citealt{DeWarf:2003aa}) and a radius of $R_*$ = 2.96~$\pm$~0.08~$R_{\odot}$ (\citealt{Stassun:2019aa}) yields an average
mass accretion rate of  log($\dot{M}$/($M_{\odot}$ yr$^{-1}$) ) = --6.28 $\pm$ 0.12 and a range of --6.3 to --6.2.
These values are similar to previous measurements for AB Aur: \citet{GarciaLopez:2006aa} found 
log($L_\mathrm{acc}$/$L_{\odot}$) = 0.63  and log($\dot{M}$/($M_{\odot}$ yr$^{-1}$)) = --6.85,
\citet{Wichittanakom:2020aa} found log($L_\mathrm{acc}$/$L_{\odot}$) = 1.32  and log($\dot{M}$/($M_{\odot}$ yr$^{-1}$)) = --6.13 , and 
\citet{Salyk:2013je} found log($L_\mathrm{acc}$/$L_{\odot}$) = 0.5  and log($\dot{M}$/($M_{\odot}$ yr$^{-1}$)) = --6.90.

Under the assumption we are actually witnessing accretion from a young planet, a mass accretion rate can similarly be 
derived for AB Aur b.  The extinction-corrected H$\alpha$ line luminosity of AB Aur b is mapped to accretion luminosity 
and mass accretion rate
using two empirical correlations: the $L_{H\alpha}$--$L_{\text{acc}}$ relation of low-mass stars and brown dwarfs from 
\cite{Alcala:2017aa}\footnote{Note that \citet{Aoyama:2021ku} propose an alternative theoretical 
$L_\mathrm{H \alpha}$-$L_\mathrm{acc}$ relation that may be more appropriate for embedded protoplanets 
experiencing accretion shocks.  They predict a higher $L_\mathrm{acc}$ for
a given $L_\mathrm{H \alpha}$, which would imply a dramatically higher accretion luminosity and accretion rate for AB Aur b.} and the observed $\dot{M}$--$M$ relation of stellar and substellar objects 
identified by \citet{Betti:2023bb}.  A radius of 2~$\Rjup$ is assumed, which corresponds to the 
typical size of a giant planet or low-mass brown dwarf to within a factor of $\sim$1.5 at young ages 
($\lesssim$5~Myr; e.g., \citealt{Phillips:2020zz}).  
Using this value for the radius together with $L_{\text{acc}}$ yields an expression for $\dot{M}$ as a function of mass $M$:
$\dot{M}$ = $R L_{\text{acc}}$/($G  M$).  
The intersection of this relation with the stellar/substellar $\dot{M}$--$M$ relation can be used to estimate
a mass accretion rate and companion mass assuming AB Aur b follows this trend.   
Results are presented in Table~\ref{tab:acclum}; the average implied accretion luminosity
is log($L_\mathrm{acc}$/$L_{\odot}$,) = --3.25 and the average accretion rate is log($\dot{M}$/($M_{\odot}$ yr$^{-1}$)) = --11.0.
The average value of log($M \dot{M}$ / ($\Msun^2$ yr$^{-1}$) is --12.6.
The corresponding average companion mass is 34~\Mjup, which is higher than estimates of 9--12~\Mjup \ for AB Aur b from \citet{Currie:2022dd}.
Note that because these are not direct measurements of the bolometric accretion luminosity, and because the host star likely contributes to the observed $F656N$ flux 
from AB Aur b (see Section~\ref{sec:scatter}),
these values should be treated with caution.

\begin{deluxetable*}{cccccc}
\renewcommand\arraystretch{0.9}
\tabletypesize{\footnotesize}
\setlength{ \tabcolsep } {.1cm}
\tablewidth{0pt}
\tablecolumns{6}
\tablecaption{Extinction-Corrected Photometry, H$\alpha$ Line Luminosity, and Mass Accretion Rate\label{tab:acclum}}
\tablehead{
    \colhead{Epoch}    &  \colhead{EC $f_{\lambda}$\tablenotemark{a}}                  & \colhead{$F_{H \alpha}$}    & \colhead{log $L_\mathrm{H \alpha}$}   & \colhead{log $L_\mathrm{acc}$\tablenotemark{b}}  & \colhead{log $\dot{M}$\tablenotemark{c}}    \\ [-0.2cm]
          \colhead{ID}       &  \colhead{(10$^{-15}$ erg s$^{-1}$ cm$^{-2}$ \AA$^{-1}$)}  & \colhead{(10$^{-10}$ erg s$^{-1}$ cm$^{-2}$)} & \colhead{($L_{\odot}$)}    & \colhead{($L_{\odot}$)}   &  \colhead{($M_{\odot}$ yr$^{-1}$)} 
        }
\startdata
 \multicolumn{6}{c}{AB Aur} \\
  \cline{1-6} 
       E0   &                           10490 $\pm$ 170   &                                       1.48 $\pm$ 0.03   &         --0.95 $\pm$ 0.01   &               1.2 $\pm$ 0.3    &             --6.2 $\pm$ 0.3    \\
       E1   &                              8516 $\pm$ 5   &                                    1.134 $\pm$ 0.0009   &        --1.064 $\pm$ 0.007   &               1.1 $\pm$ 0.3    &             --6.3 $\pm$ 0.3    \\
       E2   &                             8655 $\pm$ 14   &                                     1.159 $\pm$ 0.003   &         --1.055 $\pm$ 0.007   &               1.1 $\pm$ 0.3    &             --6.3 $\pm$ 0.3    \\
       E3   &                             7884 $\pm$ 17   &                                     1.023 $\pm$ 0.003   &         --1.109 $\pm$ 0.007   &               1.1 $\pm$ 0.3    &             --6.3 $\pm$ 0.3    \\
       E4   &                              8895 $\pm$ 9   &                                     1.201 $\pm$ 0.002   &        --1.039 $\pm$ 0.007   &               1.1 $\pm$ 0.3    &             --6.3 $\pm$ 0.3    \\
       E5   &                             7727 $\pm$ 13   &                                     0.996 $\pm$ 0.002   &         --1.121 $\pm$ 0.007   &               1.1 $\pm$ 0.3    &             --6.3 $\pm$ 0.3    \\
  \cline{1-6} 
\multicolumn{6}{c}{AB Aur b\tablenotemark{d}}  \\
  \cline{1-6} 
       E0   &                             2.3 $\pm$ 0.3   &                      4.0 $\pm$ 0.5 $\times$ 10$^{-4}$   &                           --4.51 $\pm$ 0.05  &   --3.35 $\pm$ 0.06   & --11.1 $\pm$ 0.7  \\
       E1   &                             4.3 $\pm$ 0.2   &                      7.5 $\pm$ 0.4 $\times$ 10$^{-4}$   &                           --4.24 $\pm$ 0.02   &  --3.05 $\pm$ 0.02   & --10.9 $\pm$ 0.7  \\
       E2   &                             1.5 $\pm$ 0.2   &                      2.7 $\pm$ 0.4 $\times$ 10$^{-4}$   &                           --4.68 $\pm$ 0.06  &   --3.55 $\pm$ 0.06  &  --11.2 $\pm$ 0.7 \\
       E3   &                             1.5 $\pm$ 0.2   &                      2.7 $\pm$ 0.3 $\times$ 10$^{-4}$   &                           --4.69 $\pm$ 0.04  &   --3.56 $\pm$ 0.05  &  --11.3 $\pm$ 0.7 \\
       E4   &                             6.5 $\pm$ 0.4   &                     11.4 $\pm$ 0.8 $\times$ 10$^{-4}$   &                           --4.06 $\pm$ 0.03  &  --2.85 $\pm$ 0.03  &   --10.8 $\pm$ 0.7 \\
       E5   &                             3.5 $\pm$ 0.2   &                      6.2 $\pm$ 0.4 $\times$ 10$^{-4}$   &                           --4.33 $\pm$ 0.03  &   --3.15 $\pm$ 0.03  &  --11.0 $\pm$ 0.7  \\
\enddata
\tablenotetext{a}{EC refers to extinction-corrected values assuming $A_V$=0.51~mag (\citealt{McJunkin:2014aa}).}
\tablenotetext{b}{For AB Aur, the $L_\mathrm{H \alpha}$-$L_\mathrm{acc}$ relation from \citet{Mendigutia:2011aa} is used 
to derive accretion luminosity. 
For AB Aur b, $L_\mathrm{acc}$ is derived from $L_\mathrm{H \alpha}$ following \citet{Alcala:2017aa}.}
\tablenotetext{c}{For AB Aur, mass accretion rate is derived assuming a stellar mass of 2.4~$\pm$~0.2~$M_{\odot}$ from \citet{DeWarf:2003aa} and a radius of 2.96~$\pm$~0.08~$R_{\odot}$ from \citet{Stassun:2019aa}. 
For AB Aur b, mass accretion rate is inferred from the stellar and substellar $\dot{M}$-$M_*$ trend from \citet{Betti:2023bb}.}
\tablenotetext{d}{We emphasize that there is considerable uncertainty associated with our inferred values of $L_\mathrm{acc}$ and $\dot{M}$ for AB Aur b.  See Section~\ref{sec:mdot} for details.}
\end{deluxetable*}

\subsection{Both Accretion Luminosity and Scattered Light?}{\label{sec:scatter}}

\citet{Zhou:2023di} found that the UV through red-optical emission from AB Aur b shows the strong
Balmer break and is therefore dominated by reflected light from its host star.  This implies that
our $F656N$ photometry from AB Aur b may also include a significant contribution of underlying photospheric and H$\alpha$ emission from
AB Aur.  In this scenario, some fraction of reflected light originates from the host, which is combined with strong variability produced by an accreting planet flickering in H$\alpha$.

We can estimate this contribution of the host star by comparing the $F645N$ continuum photometry of AB Aur and AB Aur b from \citet{Zhou:2023di}.
The corresponding brightness ratio is 0.00032. If we apply this to the extinction-corrected H$\alpha$ photometry of AB Aur so as to scale it
down to the expected reflected-light H$\alpha$ intensity, the resulting flux densities are broadly similar to the observed values for AB Aur b.
In particular, the ratio between the observed flux density of AB Aur b and the scaled down brightness of AB Aur
is 0.76 $\pm$ 0.10, 1.75 $\pm$ 0.08, 0.60 $\pm$ 0.08, 0.66 $\pm$ 0.09, 2.5 $\pm$ 0.16, and 1.6 $\pm$ 0.09
for epochs E0 through E5, respectively.
In this scenario, values above 1.0 would indicate that the $F656N$ emission from AB Aur b is dominated by emission from the companion,
while values near 1.0 imply that the emission is primarily reflected light from the host star.  Values below 1.0 imply that the
observed flux is less than the scaled host star contribution, probably a sign that the scaling factor based on the $F645N$-band
observations is not constant over time.  It could also indicate extinction is severe or variable.
These ratios appear to fluctuate about 1, with the highest value of 2.5 corresponding to Epoch 4.  
This could be a result of real variations from AB Aur b, whereby 
stochastic accretion sometimes elevates the source brightness substantially above the reflected light contribution.

\subsection{Disk Features and Limits on H$\alpha$ Sources}

These sensitive HST observations offer an opportunity to search for faint signals in H$\alpha$ associated with AB Aur, including
scattered light disk features and young accreting planets.
Figure~\ref{fig:coadded} shows the coadded image of all 6 epochs of observations taken in \emph{F656N} (E0--E5),
which represents 2.5 hours of total integration time across 25 orbits.
As far as we are aware, this is the deepest high-contrast imaging of AB Aur in H$\alpha$ to date. 
Although these combined individual reductions are not optimized to bring out extended emission and minimize 
self subtraction (e.g., \citealt{Mazoyer:2020aa}; \citealt{Flasseur:2021aa}; \citealt{Pairet:2021aa}; \citealt{Juillard:2023xt}), 
many known structures associated with the disk are evident.
For consistency with previous work in this bandpass, we adopt the same naming convention for the spiral arm features
from \citet{Zhou:2022fa} with subscript $H \alpha$ appended to avoid ambiguity 
with other spiral arm features on larger scales (e.g., \citealt{Hashimoto:2011bt}; \citealt{Speedie:2024aa}).

The most prominent structures are $S1_{H \alpha}$, $S2_{H \alpha}$, $S3_{H \alpha}$, and $S4_{H \alpha}$, 
each of which span several hundred AU in scale
and radiate outward in the same direction---clockwise at increasing separations.
In addition, we also identify the persistent structures $S5_{H \alpha}$ and $S6_{H \alpha}$ which are visible in this image
but were not significant in previous shallower H$\alpha$ observations.
The brightest of these features are also visible in HST UV-through-optical imaging of the system
from \citet{Zhou:2023di}, and high-resolution VLT/SPHERE near-infrared 
imaging (e.g., \citealt{Boccaletti:2020ba}; \citealt{Currie:2022dd}).
The orientation of the spiral arms are consistent with the disk kinematics from \citet{Tang:2017gr};
the disk is rotating in the counterclockwise direction and the spiral arms are trailing
the bulk motion of the gas disk.

No point-like H$\alpha$ sources are evident within or immediately outside the 
scattered light disk ($<2$'').
There have been several planet candidates noted within the AB Aur disk, but many of these are within the
effective inner working angle of $\approx$0$\farcs$4 in our observations where contrasts are poor and PSF subtraction 
is unreliable.  In particular, \citet{Boccaletti:2020ba} identified two compact sources at sub-arcsecond separations
in polarized and unpolarized $H$-band observations with VLT/SPHERE.  Source \emph{f1} is located  0$\farcs$16
south of AB Aur and is associated with extended emission and a twist in the inner spiral structure at tens of AU, which may be related to interactions 
 between a young planet and the inner disk.  \emph{f1} is too close to AB Aur to reliably search for excess H$\alpha$ emission
 in these datasets.
 
 The point source \emph{f2} is located 0$\farcs$681 north of AB Aur near the inner edge of the transition disk.
We do not recover \emph{f2} in H$\alpha$, although the neighboring spiral feature just north of \emph{f2} 
that is present in the \citet{Boccaletti:2020ba} observations is also recovered in our HST observations 
($S3_{H \alpha}$).  Similarly, hints of
the curved, trailing tip of their ``S1'' inner spiral are visible in $H\alpha$ immediately south-east of the expected position of \emph{f2}.
Both \emph{f1} and \emph{f2} are detected in polarized-intensity observations from \citet{Boccaletti:2020ba};
as a result, they conclude that these sources most likely correspond to scattering from dust in the disk.  
The morphology of \emph{f1} and its relationship to the inner spiral arms are most closely connected to expectations
of an embedded young planet, whereas \emph{f2} may instead be structure within the disk.
If \emph{f2} is actually a protoplanet, the lack of H$\alpha$ emission indicates that it is embedded with a high extinction
or it is not accreting.

No H$\alpha$ sources are evident at wider separations.  The 3$\sigma$ upper limit 
in an annulus from 1$\farcs$75--2$\farcs$56 (270--400 AU)
is 0.046 e$^-$ s$^{-1}$, or $f_{\nu}$ = 0.0033 mJy ($f_\lambda$ = 2.3$\times$10$^{-18}$ erg s$^{-1}$ cm$^{-2}$ \AA$^{-1}$).
Assuming a reddening of $A_V$=0.51~mag to AB Aur (\citealt{McJunkin:2014aa}), the upper limit corrected for extinction is 
$f_\lambda$ = 3.2$\times$10$^{-18}$ erg s$^{-1}$ cm$^{-2}$ \AA$^{-1}$.
Using the \emph{F656N} filter width of $\Delta \lambda$ = 17.6 \AA, this corresponds to an upper limit on the H$\alpha$ line flux of 
5.7$\times$10$^{-17}$ erg s$^{-1}$ cm$^{-2}$.  At the distance to AB Aur (155.9~pc), the limit on the line luminosity is 
1.7$\times$10$^{26}$ erg s$^{-1}$,
or  4.3$\times$10$^{-8}$ $L_{\odot}$.

\section{Discussion}{\label{sec:discussion}}

The results of this accretion light echo experiment offer two main insights about AB Aur b:
the scenario of stable scattering of H$\alpha$ emission from the host star is not supported by our HST observations (Figure~\ref{fig:onetoone}), 
and large-amplitude variability is occurring over timescales of weeks to years.
This implies that the variability originates very close to either the star or the companion where dynamical timescales are short.
Below we discuss these interpretations in more detail.  In Section~\ref{sec:varcomp} we compare the variability of AB Aur b with
what is known about the timescales and amplitudes of accretion in the planetary domain.
Alternative interpretations unrelated to an accreting planet are discussed in 
Section~\ref{sec:interpretations}.

\subsection{How Common is AB Aur b-like Variability?}{\label{sec:varcomp}}

The factor of four variation in brightness from AB Aur b implies 
a proportional change in the H$\alpha$ line flux and---if it is indeed a protoplanet--- accretion luminosity,
assuming  the same fraction of $L_\mathrm{acc}$ emerges in H$\alpha$ 
across all epochs of this program.
How typical is this scale of variability compared to other accreting planets, and can this lend insight into the nature of AB Aur b?
Here we contrast the high level of H$\alpha$ variability of AB Aur b with other young planets and 
widely separated planetary-mass companions.
Although the immediate surroundings and evolutionary phase of AB Aur b differ from other well-studied systems,
this comparison nevertheless provides context for whether the H$\alpha$ variability
behavior of AB Aur b is largely consistent or discrepant with what is known 
about accretion in the planetary domain.

\subsubsection{Comparison with PDS 70 b and c}

The most prominent uncontested examples of accreting protoplanets are the young giant planets PDS 70 b and c.
The most measurements in H$\alpha$ have been made for PDS 70 b, the inner of the two imaged planets.
PDS 70 b was first imaged in H$\alpha$ by \citet{Wagner:2018bb} using Magellan/MagAO, and since then 
has been followed up by \citet{Haffert:2019ba} with VLT/MultiUnit Spectroscopic Explorer---observations that were subsequently
reanalyzed by \citealt{Hashimoto:2020gn}---and HST/WFC3 (\citealt{Zhou:2021ky}).
More recently, \citet{Close:2025aa} report multi-year observations of the PDS 70 planets in H$\alpha$ with Magellan/MagAO-X,
and Zhou et al. (2025, in press) present new HST observations of the system; these are discussed in
detail below.

\citet{Zhou:2021ky} carried out an analysis of H$\alpha$ variability of PDS 70 b up to that point.
There are some signs of modest brightness changes at 3.5$\sigma$, but they rule out significant variability
at the $>$30\% level.  They also note that the possible brightness changes could instead be caused by unaccounted-for systematic biases associated with 
the variety of instruments, observing strategies, and post-processing techniques used in these studies.

This picture for both PDS 70 b and c has recently come into focus with results from \citet{Close:2025aa} and Zhou et al. (2025, in press).
\citet{Close:2025aa} recovered PDS 70 b and c from MagAO-X observations in H$\alpha$ from 2022 to 2024.
PDS 70 b decreased by a factor of 4.6 and PDS 70 increased by a factor of 2.3 during this timeframe.
By 2024, PDS 70 c appeared brighter than b, indicating long-term trends related to evolving mass accretion.  
Zhou et al. find similar results with new HST imaging in H$\alpha$; both planets are recovered in observations 
from 2024 and a clear brightening of PDS 70 c is detected when compared with observations from 2020.
Altogether this points to variations on year-long timescales with amplitudes that are broadly similar to what we find for AB Aur b,
although the timescales of brightening and fading appear to be longer for the PDS 70 planets than for AB Aur.

It is possible that these differences might be attributed to distinct evolutionary stages of protoplanetary growth between PDS 70 b, PDS 70 c, and AB Aur b.
The inferred accretion rate of $\dot{M}$ $\approx$ 10$^{-8}$ $M_\mathrm{Jup}$ yr$^{-1}$ 
(\citealt{Wagner:2018bb}; \citealt{Haffert:2019ba}; \citealt{Hashimoto:2020gn}; \citealt{Zhou:2021ky}; \citet{Close:2025aa}; Zhou et al. 2025, in press) 
implies that PDS 70 b and c are
near the end of their primary phase of growth,
and their accretion luminosity is about 300 times lower than what was inferred for AB Aur b in Section~\ref{sec:mdot}.
However, if AB Aur b is indeed an accreting protoplanet, the extended nature of the source suggests that it may be 
embedded within AB Aur's protoplanetary disk, perhaps with an extended envelope and large circumplanetary disk.
If there is a correlation between the rate of mass accretion and level of variability, then the earlier evolutionary stage of AB Aur b compared to PDS 70 b and c
could also explain potential differences
between the observed variability properties of these two case studies.
A clearer understanding of the variability properties of protoplanets will be needed to 
map the behavior of accretion over time.

\subsubsection{Comparison with Accreting Planetary-Mass Companions}

Active accretion, hydrogen line emission, and circumplanetary disks are common among 
long-period giant planets and low-mass brown dwarf companions orbiting young stars.
\citet{Bowler:2017kt} found that at least half of widely separated ($\gtrsim$100~AU) substellar companions
with masses below $\approx$20~\Mjup \ and ages $\lesssim$15~Myr show Pa$\beta$ emission.  Owing to the wavelength dependence
of dust absorption, this near-infrared line is less sensitive to extinction 
and potential confusion with chromospheric activity compared to H$\alpha$ and implies ongoing mass accretion.
\citet{Martinez:2022aa} found a comparable lower limit on the occurrence rate of circumplanetry disks based on thermal excess emission.

How useful is this population as a comparison sample for AB Aur b?
Based on their orbital eccentricities, stellar obliquities, and atmospheric compositions, there is growing evidence that the dominant origin of wide planetary companions  is 
more aligned with gravitational fragmentation of molecular cloud cores or protoplanetary disks 
compared to core- or pebble-accretion channels typically associated with giant planets
at closer separations (e.g., \citealt{Wu:2017kd}; \citealt{Pearce:2019iv}; \citealt{Bowler:2020hk}; 
\citealt{Bryan:2021aa}; \citealt{PalmaBifani:2022aa}; \citealt{Bowler:2023aa}; \citealt{Balmer:2024aa}; \citealt{Xuan.2024inb}).
Nevertheless, despite their older ages compared to AB Aur b, 
the high prevalence of subdisks among these objects and relative ease of moderate-resolution spectroscopy 
has opened the door to regular spectroscopic monitoring 
spanning a variety of timescales.
In this sense, if AB Aur b is a young planet, it could represent an example of a younger analog of these objects.

Recently, \citet{Demars:20232dl} carried out a detailed analysis of the Pa$\beta$ emission line variability for two widely-separated companions
near the deuterium-burning limit, GQ Lup~B and GSC~0614-210~b.
They found that variability is typically modest on short timescales of hours to weeks but dramatically increases to 100--1000\% 
on timescales of years to decades.  
Other studies have came to similar conclusions in the substellar regime.
\citet{Stelzer:2007hl} identified order-of-magnitude variability on timescales of months to years from the brown dwarf host 2MASS J12073346--3932539 A.
\citet{Wu:2023mud} found that H$\alpha$ variability of the $\approx$19~\Mjup \ companion FU Tau B
is relatively low (10--20\%) on short timescales of hours to days.
\citet{Wolff:2017ky} showed that the Pa$\beta$ emission line of DH Tau B varies by a factor of a few on timescales of weeks to months. 
In the thorough compendium of accreting objects spanning stars to planets, \citet{Betti:2023bb} found that the spread of $\approx$100--300\%  
in accretion variability down to the deuterium-burning limit is not particularly unusual compared to trends among brown dwarfs and stars.

There is clearly a large dynamic range of variability amplitudes among accreting giant planets and wide planetary-mass objects.  
These broadly correlate with sampling timescales, with higher variability observed over longer time baselines.
This variability amplitude and timescale is consistent with what we find for AB Aur b, given our limited sampling.
We conclude that if the observed H$\alpha$ of AB Aur b originates from accretion onto a protoplanet, the observed variability is within the 
range of accreting wide-orbit planetary mass companions.
We also note that this variability could impact population statistics from direct imaging surveys
searching for accreting young planets using emission lines (\citealt{Plunkett:2024aa}), as 
young planets may be missed during quiescent phases of accretion.  Multiple epochs over long timescales could help overcome this possible bias
in protoplanet demographics.

\subsection{Alternative Interpretations}{\label{sec:interpretations}}

\subsubsection{Variable Extinction?}

Changes in the H$\alpha$ line luminosity, accretion luminosity, and mass accretion rate 
discussed in Section~\ref{sec:mdot}
inherently assume the observed variations relate to stochastic accretion.
However, the H$\alpha$ variability from AB Aur b could alternatively be caused by evolving line-of-sight extinction 
from accretion flows, circumplanetary material, or circumstellar material,
altering what may actually be a more constant mass accretion rate onto  
a young planet (e.g., \citealt{Marleau:2022bb}).  

The circumplanetary environment surrounding a 
protoplanet is expected to be complex, but there are few observational constraints to help anchor the picture of
giant planet growth at the earliest embedded stages.
The immediate region near a growing planet is likely to include a circumplanetary disk out to an appreciable
fraction of the Hill radius (e.g., \citealt{Ward:2010fc}; \citealt{Fung:2019aa}; \citealt{Taylor:2024aa}; \citealt{Zhu:2015fr}).  
Prior to a gap-opening phase when circumstellar disk surface densities diminish, 
the circumplanetary disk may take the form of a decretion disk with gas circulating via meridional flows and 
accretion proceeding along the polar regions (\citealt{Tanigawa:2012jk}; \citealt{Batygin:2018hq}; \citealt{Szulagyi:2022aa}).
Magnetically driven outflows from a young planet might accompany this 
process (\citealt{Quillen:1998dw}; \citealt{Fendt:2003fo}; \citealt{Law:2023vd}; \citealt{Yoshida:2024aa}).
All of this can give rise to variable extinction to the growing planet.

What are the implied extinction values given the observed variability of AB Aur b?  Taking the brightest 
epoch (E4) of AB Aur b as a reference point, we can deredden the other measurements of AB Aur b at
other epochs until we reach this value.  The resulting extinctions are $\Delta A_V$ = 1.5~mag, 0.6~mag, 2.1~mag, 2.1~mag, and 0.9~mag
for epochs E0, E1, E2, E3, and E5, respectively, using the reddening law from \citet{Fitzpatrick:1999dx}.
Given the timing of the observations, with E1--E3 taken within
a two-month period in 2023 and E4 and E5 taken within a two-week period in 2024, this would indicate
that changes in extinction at the scale of 1--2~magnitudes is occurring on timescales of weeks to months.

It is unclear whether these constraints on the timescale and amplitude of variations 
alone can be used as a meaningful test of 
this variable extinction hypothesis because of the potential complexity
of the circumplanetary region.  Each physical process will operate on its own characteristic timescale---Keplerian
motion of a disk about the planet, high-velocity outflows expelling gas and entraining dust, accretion 
and meridional flows, and the planetary rotation rate.  Some 
other relevant timescales are discussed in \citet{Chiang:1997ev}.
Of course, it is also possible that both extinction and changes in accretion luminosity could both be taking 
place simultaneously, which would be challenging to disassociate.

\subsubsection{Inner Disk Shadowing?}

The dissimilar variability between AB Aur A and b has thus far been discussed in the context of AB Aur b as an accreting planet.
However, there are other plausible, and perhaps even likely, explanations for the inconsistent 
variability between AB Aur and AB Aur b that are unrelated to emission from a young planet.
These alternative interpretations should be considered alongside the planet hypothesis.

One possibility is that AB Aur b is a compact disk feature seen in scattered light and that 
brightness variations are caused not by the source itself, but from structural changes in the inner disk located along the 
sight line from AB Aur b to its host star.  This type of inner disk shadowing is becoming increasingly recognized
as a common phenomenon among protoplanetary disks with dynamic inner regions.
Examples of shadowing structures include inner disk misalignments 
(e.g., \citealt{Benisty:2018cc}; \citealt{Pinilla:2018aa}; \citealt{Zhu:2018aa}; \citealt{Bohn:2022aa}; 
\citealt{Debes:2023tv}; \citealt{Villenave:2024aa}), warps (\citealt{Nealon:2019aa}; \citealt{Kluska:2020aa}), 
``puffed up'' inner disk walls with variable scale heights (\citealt{Sitko:2008aa}; \citealt{Wisniewski:2008aa}; \citealt{Muzerolle:2009ez}),
dust transport caused by magnetic activity close to the star (\citealt{Turner:2010aa}),
and dust entrained in outflows from the inner disk (\citealt{Liffman:2020aa}).
Some of this inner disk evolution is expected to be connected to orbiting planets, which can perturb the nearby disk
structure not just in the disk's midplane but also in its vertical structure.  Indeed, several planet candidates
have been identified in the inner regions of the AB Aur disk (\citealt{Boccaletti:2020ba}) which could be related to the observed
variability of AB Aur b if they can impact the disk structure at sub-AU scales.  

What adds more credence to this picture is that the AB Aur disk has itself been shown to exhibit 
brightness variations.  \citet{Shenavrin:2019aa} found in unresolved imaging observations 
that AB Aur is variable by up to 0.7~mag from 1--5~$\mu$m where inner disk emission 
dominates over the stellar photosphere (see also \citealt{Shenavrin:2012aa}).  
At longer wavelengths, \citet{Prusti:1994aa} identified brightness changes 
of 0.27~mag at 12~$\mu$m and 0.17~mag at 25~$\mu$m with IRAS over a 6-month period.  
\citet{Chen:2003aa} measured comparable variability at the 20--50\% level at 11.7~$\mu$m and 18.7~$\mu$m
from observations at Keck Obesrvatory.
This temporal variability is suggestive of changing disk illumination, although the origin and specific nature of the variability is unclear given 
the limited long-wavelength monitoring of AB Aur.

An outer limit of the location of shadowing can be estimated from the minimum timescale over which variability is observed.
For this dataset, this occurs between Epochs 4 and 5, which are separated by about 12 days.  Assuming a stellar mass of 2.4~\Msun,
this corresponds to an orbital distance of 0.14~AU.  Physical variations of the inner disk are expected at smaller spatial scales,
so in this scenario shadowing is likely to originate closer in.  Moreover, this estimate is limited by sampling cadence; shorter timescale variability 
would imply vertical structure at even shorter orbital periods.

The inner regions of AB Aur's disk have been probed at high spatial resolution with optical and near-infrared interferometry
(e.g., \citealt{MillanGabet:2001aa}; \citealt{Eisner:2004aa}; \citealt{Millan-Gabet:2006aa}; \citealt{Tannirkulam:2008aa}).  
In a comprehensive modeling effort by \citet{Tannirkulam:2008jv}, they found an inner dust disk with an inner radius of 
0.24~AU, which is similar to spatial scales of resolved structures identified with interferometry.
The dynamical (Keplerian) timescale at this separation is of order a month, and any inhomogeneities of the inner disk
could result in changes to shadowing on much shorter times of only a fraction of the orbital period.  
This is broadly consistent with our observations of AB Aur b, suggesting that inner shadowing of inner disk
structure could plausibly explain the observed variability.

Finally, there are also signs that the AB Aur disk exhibits a global warp based on inclination measurements 
from small separations out to large spatial scales corresponding to the outer disk
(\citealt{Hashimoto:2011bt}; \citealt{Tang:2012tt}; \citealt{Speedie:2024aa}).  
The origin of the warp is unclear, but may be related to a close-in inclined planet 
similar to behavior in the $\beta$~Pic disk (e.g., \citealt{Mouillet:1997ib}; \citealt{Apai:2015aa}).
Disk warps can also result from the dynamical influence on an outer companion.  AB Aur is a wide binary with the nearby 
disk-bearing star SU Aur (e.g. \citealt{Herbig:1960aa}; \citealt{Eggen:1975aa}; \citealt{DeWarf:2003aa}), 
but the projected separation of 26,800~AU is likely too large to significantly impact AB Aur's disk\footnote{This physical 
connection between AB Aur and SU Aur is strengthened with Gaia DR3 astrometry (\citealt{GaiaCollaboration:2022aa}).  AB Aur has a parallax of 6.413 $\pm$ 0.037~mas and proper motions of $\mu_{\alpha} \cos \delta $ = 4.02 $\pm$ 0.04~mas yr$^{-1}$ and $\mu_{\delta}$ = --24.03 $\pm$ 0.04~mas yr$^{-1}$.  The parallax of SU Aur is 6.370 $\pm$ 0.030~mas and proper motions of $\mu_{\alpha} \cos \delta $ = 4.19 $\pm$ 0.03~mas yr$^{-1}$ and $\mu_{\delta}$ = --24.30 $\pm$ 0.02~mas yr$^{-1}$.
Both stars have slightly elevated Gaia Renormalized Unit Weight Error (RUWE) values of 1.37 for AB Aur and 1.51 for SU Aur, which may be suggestive of a 
close-in companion (\citealt{Stassun:2021hr}), although \citet{Fitton:2022aa} propose a higher RUWE threshold of 2.5 for young disk-bearing stars as a result of 
excess astrometric variability in these systems.}.
Regardless, precession of a warped inner disk can also contribute to variability of the outer disk
and could be connected to the observed H$\alpha$ brightness changes.

\subsubsection{A Dynamic Disk Feature?}

Another way in which a compact disk feature could result in variability is if the structure itself is evolving,
and in the process changing the effective scattered light surface area along our line of sight.
An overdensity in the disk can experience a change in morphology through a range of processes including shearing as a result of orbital motion or
heating from exposure to stellar irradiation.  
There are many examples of evolving small-scale disk features, which are often located at small spatial scales in 
regions vulnerable to heating from the star (\citealt{Huelamo:2011hx}; \citealt{Olofsson:2011ky}; 
\citealt{Kraus:2012gk}; \citealt{Cheetham:2015hg}; \citealt{Sallum:2015gm};  \citealt{Sallum:2015ej}; \citealt{Follette:2017jw}; \citealt{Gratton:2019aa}).
Recently, \citet{Sallum:2023aa} found evidence of small-scale structures in the LkCa 15 disk at $\sim$20~AU that are
undergoing both brightness and morphological changes on timescales of years.
This may be what we are witnessing with AB Aur.

\subsection{Testing These Scenarios}

Here we comment on a few observational tests that might be used to distinguish these scenarios.
Differential emission line profiles and velocities between AB Aur A and b would be a telltale signature of a planet,
as the freefall velocity of accreting gas falling onto a massive star 
should be substantially broader than the velocity wings for a planet.  
This approach was used by \citet{Haffert:2019ba} to argue that PDS 70~b and c
are both accreting planets based on their H$\alpha$ emission line widths, shapes, 
and velocity offsets.
Like PDS 70, the H$\alpha$ emission line from AB Aur shows an inverse P Cygni profile,
and in principle could be used to discriminate these scenarios.
However, this is complicated by the scattered-light contribution of AB Aur,
which may be substantial (Section~\ref{sec:scatter}), and the wide orbital distance implies a 
velocity offset of only a few km s$^{-1}$.

Another approach to distinguish these scenarios is to simultaneously monitor AB Aur b variability in the H$\alpha$ 
line and a neighboring continuum region.  Changing shadows or a dynamic disk feature should impact both 
continuum and line emission in the same way, whereas a constant continuum but changing H$\alpha$ line strength 
would point to an accreting plant.  
Furthermore, accretion luminosity and extinction can potentially be separated through
simultaneous monitoring of multiple accretion tracers in the near-infrared where extinction is lower, 
for instance with Paschen or Brackett series recombination 
emission lines (e.g., \citealt{Biddle:2024ni}; \citealt{Marleau:2024vw}).
Variable extinction would alter line ratios as a function of wavelength, 
while variable mass accretion would impact overall line intensities.

Tracking H$\alpha$ brightness variations at other positions within the AB Aur scattered light disk can also be used
to determine whether the observed variability from AB Aur b is typical or unique to this source.
Distinct changes from AB Aur b would provide strong evidence that it is a planet if the
rest of the disk varies in sync.  
However, a careful accounting of the light echo effect at other locations is required, as 
an impulse of emission from the host star would reverberate at different times across the disk.  
It would also require careful forward modeling of the disk to avoid the impact of self-subtraction---something that may be
possible with this HST dataset but which is beyond the immediate goals of this study.

Finally, continued variability monitoring of AB Aur b over a wide range of timescales should offer further clues about
the origin of the brightness changes.  Short-cadence observations over days and weeks would sample rapid changes caused by 
mass accretion, inner disk features, and gas dynamics in the circumplanetary region.  Over months and years, longer-term
trends from slower processes connected to orbital motion of the planet or precession of a warped disk should be evident.

\subsection{Applications of Accretion Light Echoes}

In this study, we use H$\alpha$  images of the host star AB Aur  
to monitor its accretion luminosity prior to acquiring the deeper observations of the companion.
Here the source geometry and light travel time to AB Aur b is known, but in 
other instances the orbital geometry of a planet candidate may be unknown
or poorly constrained.  
In these scenarios, we recommend a long baseline of continuous H$\alpha$ monitoring of the host star leading up to and during the 
high-contrast imaging observations.
This could be carried out even with low-resolution optical spectroscopy, ideally commencing $\sim$1.4 $\times$ ($r$/(10 AU)) hours before
the deep imaging, where $r$ is the separation to the source in AU.
This scaling corresponds to the maximum light travel time to a given orbital distance.
For probing wide separations of order 100~AU, 
this may not be practical to carry out for ground-based telescopes at the same site or with the same facility because of the long light travel time.
In this case longitudinal site coverage may help---for instance,
spectroscopic monitoring from Chile prior to deep H$\alpha$ imaging from Hawai'i.
Long delay times may be more practical for space-based facilities such as HST.

In principle, an H$\alpha$ light curve of a variable host star and companion could be used not just to validate or refute a candidate planet,
but also to measure the stellocentric distances to features within a disk.
With high enough cadence and a long time baseline of the companion, the time delay measured from
a cross correlation of light curves at different positions in the disk would map disk structures in three dimensions---in essence, 
reverberation mapping of resolved disks.
Much of this will be facilitated with the next generation of 25--40-meter telescopes with sensitivity to smaller spatial scales,
fainter disk structures, and lower-mass protoplanets.

\section{Summary}{\label{sec:summary}}

Distinguishing compact disk features from accreting planets is a significant observational challenge.
We have presented a framework for using accretion light echoes to probe the nature of candidate accreting protoplanets around young stars.
Well-timed H$\alpha$ observations of the host star that take into account the light travel time to the companion offer a new approach to 
disentangle scattered light from planetary accretion luminosity through correlated brightness variations.  

We applied this method to the candidate planet AB Aur b with five epochs of high-contrast H$\alpha$ ($F656N$) imaging using HST/WFC3.
Each epoch sampled the host star's brightness prior to deep imaging of the candidate companion,
with the delay time corresponding to the observed light travel time delay to AB Aur b.
Below is a summary of our results.

\begin{itemize}
\item AB Aur b is recovered in all five epochs taken in 2023--2024.  The source is spatially resolved with an average 
aspect ratio of $\sigma_x / \sigma_y$ = 1.5.  The elongation in the azimuthal direction, 
consistent with previous results with ground- and space-based imaging.  
The corresponding deprojected size of the AB Aur b structure is 91~mas $\times$ 57~mas, or 14~AU $\times$ 9~AU.
We also reprocessed previous HST observations of AB Aur in H$\alpha$ from \citet{Zhou:2022fa} 
in a consistent fashion to compare with results from this program.
This earlier dataset did not sample the host star with a time delay, but extends H$\alpha$ photometric monitoring of AB Aur b with HST to two years.
\item AB Aur b is highly variable at the 330\% level while the host star exhibits more modest brightness changes of 15\% during this monitoring campaign.
Variations at the 100-200\% level from AB Aur b are observed on timescales of weeks to months.  
This level of variability is not unusual for known accreting planetary-mass companions and free-floating giant planets on comparably long timescales.  
\item The variability between AB Aur A and b do not show strong covariance.  This disagrees with expectations if AB Aur b is purely unobstructed, passive scattered emission.
These results either bolsters evidence that AB Aur b is an accreting planet, or we are witnessing variations associated with a dynamic disk environment.
\item Assuming the emission is entirely accretion luminosity from a protoplanet, we find an average H$\alpha$ line luminosity of log($L_{H \alpha}$/$L_{\odot}$) = --4.4.
Using the $L_{H\alpha}$--$L_{\text{acc}}$ relation from \cite{Alcala:2017aa} yields an average accretion luminosity of log($L_\mathrm{acc}$/$L_{\odot}$) = --3.25.
We caution that this depends on many unknown factors, most notably potential extinction associated with the immediate environment surrounding AB Aur b.
Nevertheless, this is over two orders of magnitude higher than PDS 70 b (\citealt{Zhou:2021ky}).  This could reflect an earlier evolutionary state 
of AB Aur b with comparably higher mass accretion.  However,  \citet{Zhou:2023di} showed that emission from AB Aur b is certainly related to a spatially extended disk feature
that is scattering emission from the host star.  When taking this into account, some epochs appear to be dominated by reflected light from AB Aur A, while at other epochs 
AB Aur b outshines the host star's contribution by up to a factor of 2.5 at maximum brightness.  
It is also possible that variable line-of-sight extinction to AB Aur b could contribute to its observed variability.  
\item If the AB Aur b feature is associated with a protoplanet then these results represent the highest level of 
H$\alpha$ variability of any imaged planet or wide planetary-mass companion to date.  To avoid missing variable signals, it would be prudent for 
direct imaging surveys using accretion signatures like H$\alpha$ to carry out multi-epoch observations 
rather than a single epoch strategy, which might happen to sample a planet during a low state of mass accretion.
\item Our coadded image of AB Aur amounts to 25 orbits, or about 2.5 hours of total integration time.  To our knowledge, this is the deepest H$\alpha$ imaging of this system to date.
AB Aur b is the most prominent source.  We recover the brightest spiral features and detect new scattered-light substructures in the disk.   No other point sources
are detected at wider separations down to 3$\sigma$ upper limits of $f_\lambda$ = 3.2$\times$10$^{-18}$ erg s$^{-1}$ cm$^{-2}$ \AA$^{-1}$.
\item There are several alternative interpretations for AB Aur b 
that should be considered alongside the accreting planet hypothesis.
It is possible that AB Aur b is simply the brightest scattered-light clump in a highly structured disk.
There are many ways to induce brightness changes in this scenario.
This includes shadowing caused by a misaligned inner disk, a puffed up inner disk wall, a disk warp, or inner disk outflows.  
Evolution of a compact dust structure can also result in brightness changes as the projected scattering surface area
changes.
Distinguishing the effects of extinction, changing illumination, inner disk scale height, and dynamic dust structures can be addressed
with multiwavelength variability monitoring over a range of timescales. 
\item More broadly, accretion light echoes may be a useful tool to distinguish scattered light disk features 
from accreting protoplanets.
Uniquely interpreting the H$\alpha$ variability can be challenging in complex disk environments like AB Aur, but this approach 
could complement other strategies to distinguish scattered-light and planetary hypotheses such as sensitive polarized intensity imaging
and differential emission line profiles.
To facilitate accretion light echo variability measurements, we recommend continuous spectroscopic 
monitoring of the host star to track accretion tracers 
prior to and throughout deep H$\alpha$ imaging observations.  
\end{itemize}

~\\ \\ 

We are grateful to the referee for their helpful feedback, 
Christopher McKee and Wenbin Lu for helpful conversations about light echoes,
and Tricia Royle at STScI for support scheduling the HST observations with the appropriate time delay for this program.
B.P.B. acknowledges support from the National Science Foundation grant AST-1909209, NASA Exoplanet Research Program grant 20-XRP20$\_$2-0119, and the Alfred P. Sloan Foundation.
This material is based upon work supported by the National Science Foundation Graduate Research Fellowship under Grant No. 2139433.
This research is based on observations made with the NASA/ESA Hubble Space Telescope obtained from the Space Telescope Science Institute, which is operated by the Association of Universities for Research in Astronomy, Inc., under NASA contract NAS 5–26555. These observations are associated with programs GO 17168 and GO 16651.
This research has made use of NASA's Astrophysics Data System Bibliographic Services

\vspace{5mm}
\facilities{HST(WFC3), TESS}

\newpage

\appendix

\section{Light Curves of AB Aur}{\label{sec:appa}}

HST undergoes heating and settling cycles when its thermal environment changes.  This can happen if HST's orientation 
relative to the Sun is altered, for instance during the course of an orbit or when slewing between targets.  The impact
of thermal changes are small focal variations that can alter the PSF shape, encircled energy in the PSF core, and photometric
measurements by a few percent  (e.g., \citealt{Anderson:2017aa}).  The scale of these systematic effects is small compared 
to the observed brightness changes of AB Aur and AB Aur b.  Here we examine whether there are signs of 
breathing patterns to determine what, if any, effect it might have on the photometry and uncertainties we have used in this work.

Section~\ref{sec:wfc3dr} describes our approach to extracting $F656N$ photometry of AB Aur in Orbit 1 of each epoch and applying an aperture correction to account
for flux loss at larger radii.  We also measure photometry of AB Aur in each frame for Orbits 2--4 of each epoch in the same fashion.  If breathing
is impacting the light curves of AB Aur, we would expect this to have the greatest impact on Orbits 1 and 2, as these data 
are taken after slews from executing other programs.

\begin{figure*}
  \vskip -3.1 in
  \hskip 0.5 in
  \resizebox{5.5in}{!}{\includegraphics{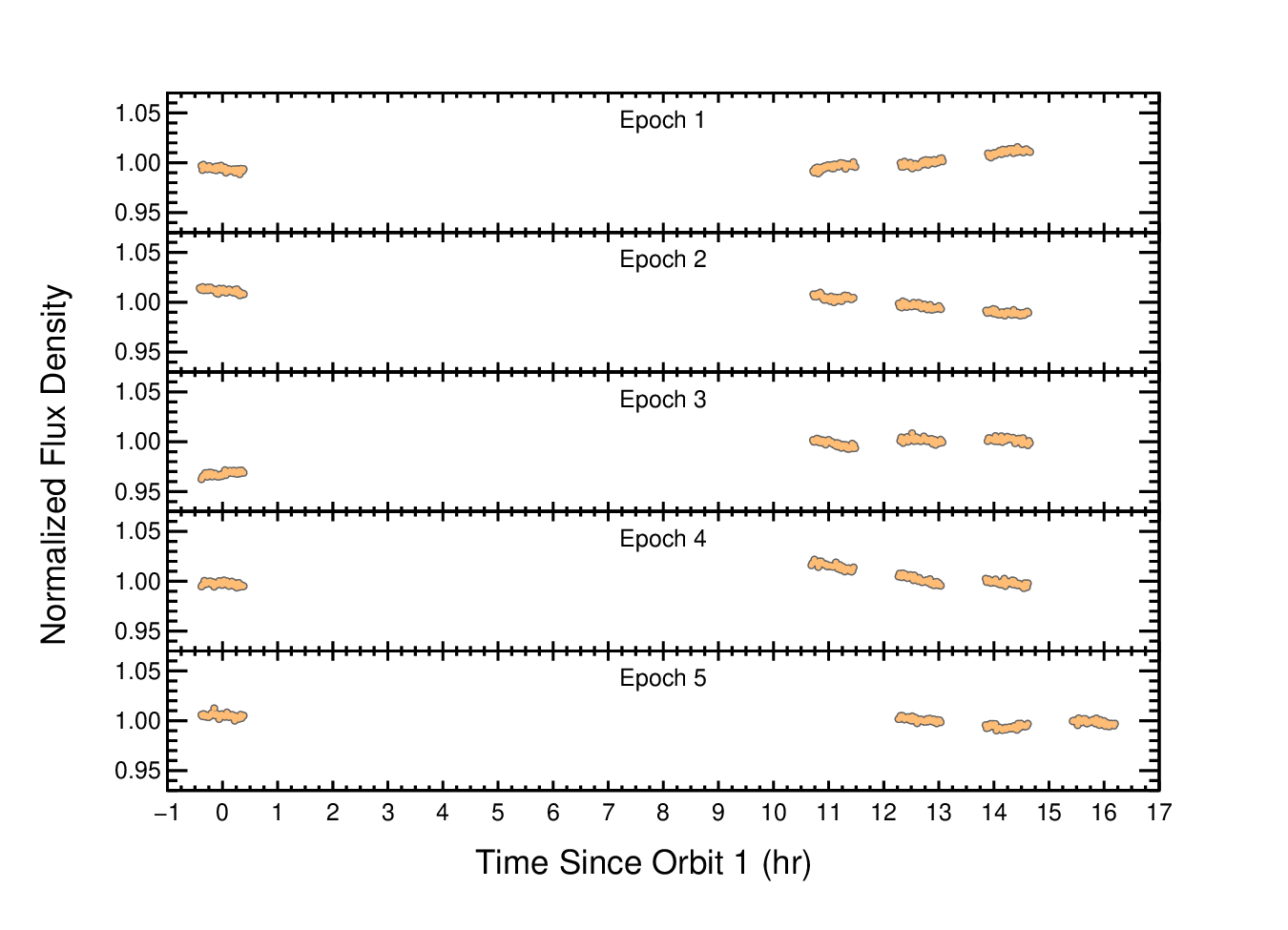}}
  \vskip -0.2 in
  \caption{Light curves of AB Aur in $F656N$ across the five epochs of this experiment.  
  Light curve timescales are with respect to the midpoint of Orbit 1, and photometry
  has been normalized to the average flux in Orbit 1.   
  Overall AB Aur is quite stable, with low-amplitude changes in H$\alpha$ at the few-percent level evident across the 15--17 hour
  baseline of each epoch.
   \label{fig:hostphot} } 
\end{figure*}

The results are displayed in Figure~\ref{fig:hostphot}.  Intra-orbit H$\alpha$ light curves are well behaved and show small, smoothly varying 
changes at the level of a few percent within a single orbit.  Inter-orbit variability is comparable, with both positive and negative trends persisting for several orbits.  The light curves undulate on few-hour timescales which is likely related to small changes in accretion rate.
We also note that photometry from Orbit 1 is in good agreement with the light curves from Orbits 2--4.  There are no signs of
of a persistent pattern in the light curve shapes from Orbits 1 and 2 that might be attributable to breathing and thermal stabilization. 


\newpage


\end{document}